\documentclass[10.5pt]{article}
\usepackage{times}
\usepackage[left=3.18cm, right=3.18cm, top=2.54cm, bottom=2.54cm]{geometry}
\usepackage{indentfirst}
\usepackage{float}
\usepackage{graphicx}
\usepackage{subfigure}
\usepackage{cite}
\usepackage[colorlinks,
linkcolor=black,
anchorcolor=black,
citecolor=black,
urlcolor=black
]{hyperref}
\usepackage{titlesec}
\titleformat{\section}{\fontsize{14}{8}\selectfont  \bfseries}{\arabic{section}}{1em}{}[]
\titleformat{\subsection}{\fontsize{12}{8}\selectfont  \bfseries}{\arabic{section}.\arabic{subsection}}{1em}{}[]
\titleformat{\subsubsection}{\fontsize{12}{8}\selectfont  \bfseries}{\arabic{section}.\arabic{subsection}.\arabic{subsubsection}}{1em}{}[]

\title{\fontsize{15}{10}\selectfont \bf{Research status and prospect of graphene materials in aviation}}
\author{\fontsize{10}{10}\selectfont Siyang Gao\textsuperscript{1,2}, Yifei Jiang\textsuperscript{1,2}, Jianwei Sun\textsuperscript{1}, Zhihui Zhang\textsuperscript{2*}\\ \fontsize{10}{10}\selectfont \textsuperscript{1}School of Mechatronic Engineering, Changchun University of Technology, Changchun 130012, China\\ \fontsize{10}{10}\selectfont \textsuperscript{2}College of Biological and Agricultural Engineering, Jilin University, Changchun 130022, China\hspace{25pt}\quad\\ \fontsize{10}{10}\selectfont Corresponding Author. E-mail: hzz314926535@126.com}
\date{}

\makeatletter
\def\@cite#1#2{\textsuperscript{[{#1\if@tempswa,#2\fi}]}}
\makeatother

\begin{document}

\maketitle 

Among various 2D materials, graphene has received extensive research attention in the past 2-30 years due to its fascinating properties. The discovery of graphene has provided a huge boost and a new dimension to materials research and nanotechnology. Many lightweight materials with good performance have been widely used in the aviation field, which has greatly promoted the development of military and civilian industries and promoted technological innovation. Based on the introduction of the structure and properties of graphene, this paper summarizes the application value of graphene in the aerospace field in three aspects: energy equipment, sensors, and composite materials used outside aircraft.

\section{Introduction}
In recent years, with the continuous development and progress of nanoscience and nanotechnology, we are pursuing new aviation materials with high reliability and long life. Graphene has a wide range of application prospects in the field of aviation materials due to its excellent mechanical, thermal, electrical, optical, tribological properties, super resistance to air permeability and super large specific surface area\cite{1}. 

\begin{figure}[h!]
    \centering
    \includegraphics[width=0.5\textwidth]{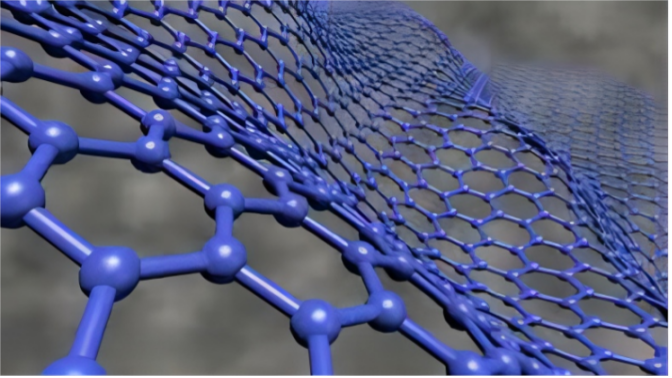}
    \caption{Wrinkles in graphene\cite{1}}
    \label{fig1}
\end{figure}

In 2004, K. S. Novoselov et al\cite{2}. obtained graphene for the first time through the micromechanical exfoliation method. It is formed by the periodic arrangement of a single layer of carbon atoms in the form of six rings in a two-dimensional(2D) plane. At the same time, there are corrugated folds of about a few nanometers on the plane, as shown in Figure \ref{fig1}.

J. M. Carlsson\cite{3} believes that there are two reasons for graphene to produce corrugated wrinkles: first, the 2D structure of graphene makes the carbon atoms on the film instability in the direction perpendicular to the graphene plane to produce wrinkles; second, carbon-carbon The bond has a certain degree of flexibility. The carbon-carbon bond length does not always maintain a constant theoretical value, but changes within the theoretical length range. Therefore, the 2D plane of graphene has certain undulations. The carbon atoms in graphene are combined by sp2 hybridization, and the carbon-carbon atoms in the 2D plane form a $\sigma$ bond, and its high robustness makes the theoretical value of graphene elastic modulus reach 1.1 TPa\cite{4}. In the direction perpendicular to the 2D plane of graphene, $\pi$ bonds with weaker binding energy are formed, and its semi-filled structure makes the electron conduction rate as high as 8$\times$10$^5$ m/s\cite{5}. At the same time, the weak $\pi$ bond also enables graphene to have a smaller shear force and thus has excellent tribological properties. In addition, graphene also has a very high specific surface area\cite{6}, thermal conductivity\cite{7}, light transmittance and gas permeability resistance and other excellent properties. Therefore, graphene meets the high-performance requirements of new aviation materials and has broad application prospects in the field of aviation materials. For example, by taking advantage of the excellent mechanical properties of graphene and adding it to resins and metals, a lightweight, high-load aviation composite material can be obtained; the high light transmittance of graphene can be used in the field of aviation solar cells; Its excellent friction properties make it hopeful to become a new type of aviation lubricating material; the preparation of graphene sensors utilizes its super large specific surface area.

Table 1: summarizes the performance characteristics of graphene and its application prospects in the aerospace field.
\begin{table}[h]
\small
\newcommand{\tabincell}[2]{\begin{tabular}{@{}#1@{}}#2\end{tabular}}
    \centering
    \begin{tabular}{ccc}
    \hline
     Performance characteristics  & Advantage & Application prospects in aviation  \\ \hline
       thinnest, hardest, strongest  & \tabincell{l}{(1) The thickness is only the thickness  \\of a single layer of carbon atoms;\\(2) Mohs hardness is higher than diamond;\\(3) Micro-strength can reach 125 GPa, \\which is more than 100 times that \\of traditional steel.}
&\tabincell{l}{Ultra-thin and ultra-light aviation \\composite material.}\\\hline
Ultrahigh electronic conductivity & \tabincell{l}{(1)2$\times$ 10 $^5$ cm $^2$/(V*s) at room temperature,\\which is 100 times that of silicon.\\(2) The theoretical value is 1$\times$106cm$^2$/(V*s), \\reaching 1/300 of the speed of light}& \tabincell{l}{Aerospace thermoelectric materials, \\integrated circuits}\\\hline
Large specific surface area&\tabincell{l}{Up to 2630m$^2$/g, much higher than \\activated carbon (1500m$^2$/g)}&Aviation gas sensor\\\hline
\tabincell{l}{High light transmittance, \\high toughness}&\tabincell{l}{The light transmittance reaches 97.7\%, \\and it will not break when stretched by 20\%}&Solar battery\\\hline
    \end{tabular}
    \label{tab1}
\end{table}
\section{Application of graphene in energy devices in the aviation field}
In aviation energy devices, graphene is mainly used in supercapacitors and solar cells. Therefore, it will have a revolutionary impact on the development of aviation industry. A supercapacitor is an energy storage device that has high power and long cycle life, does not cause excessive pollution to the environment, and can achieve rapid charging and discharging functions. When graphene is used as the electrode of a supercapacitor, the specific capacity is about 100mAh/g. In the aviation field, supercapacitors can be widely used. For example, when the aircraft opens the door, it needs a particularly strong explosive power, and the supercapacitor will complete this function. It provides explosive power for the aircraft to open the door, with a service life of 25 years and 140,000 flight hours, which has been certified by Airbus. Solar cells can convert solar energy into electrical energy. After combining with graphene, it can significantly improve the transparency and conductivity of the material through doping or modification\cite{8}. Graphene solar cells are mounted on drone wings to extend drone flight time. Graphene material is an important opportunity for the development of electronic components in the aerospace field.
\subsection{Supercapacitors and graphene}
\begin{figure}[h!]
\centering
\subfigure[]{\label{fig2a}
\includegraphics[width=0.35\linewidth]{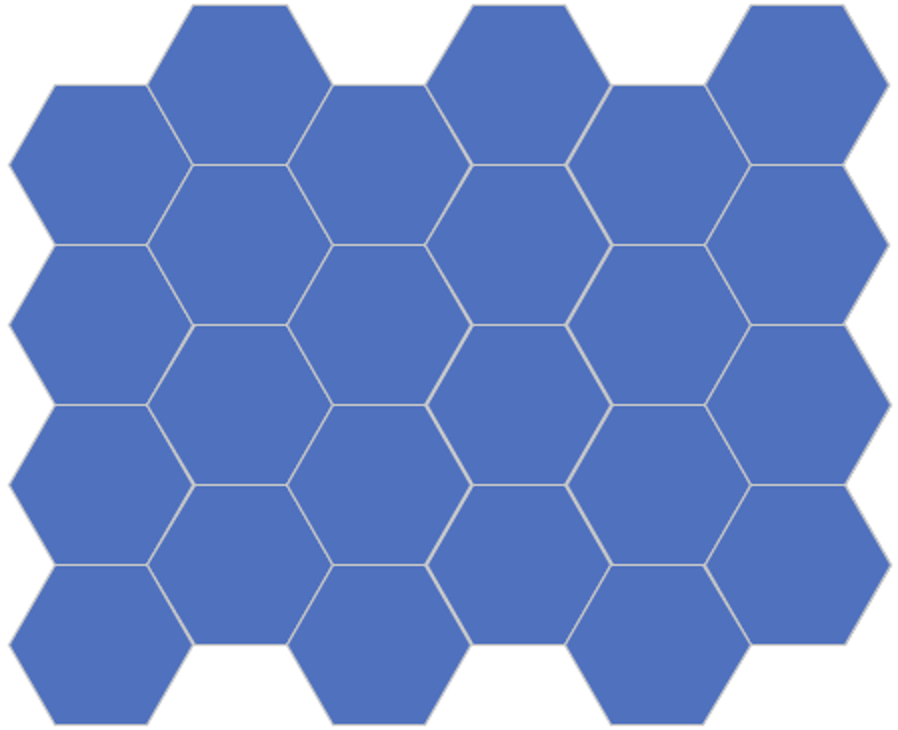}}
\hspace{0.01\linewidth}
\subfigure[]{\label{fig2b}
\includegraphics[width=0.5\linewidth]{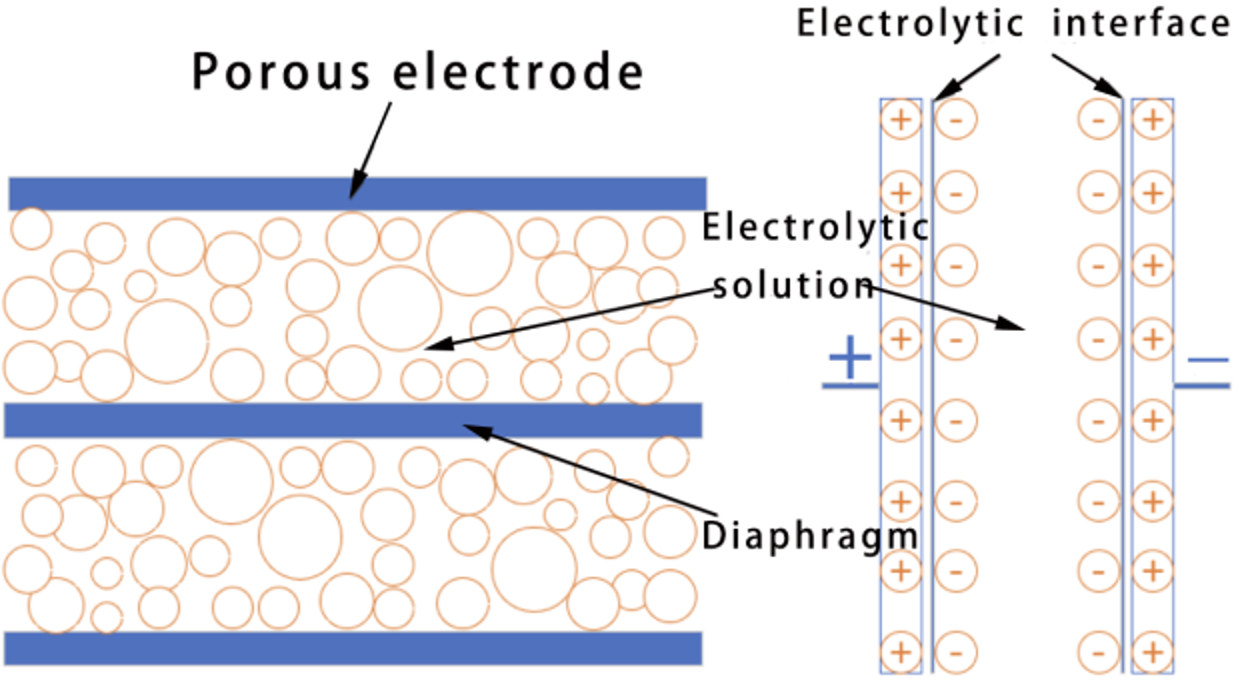}}

\subfigure[]{\label{fig2c}
\includegraphics[width=0.5\linewidth]{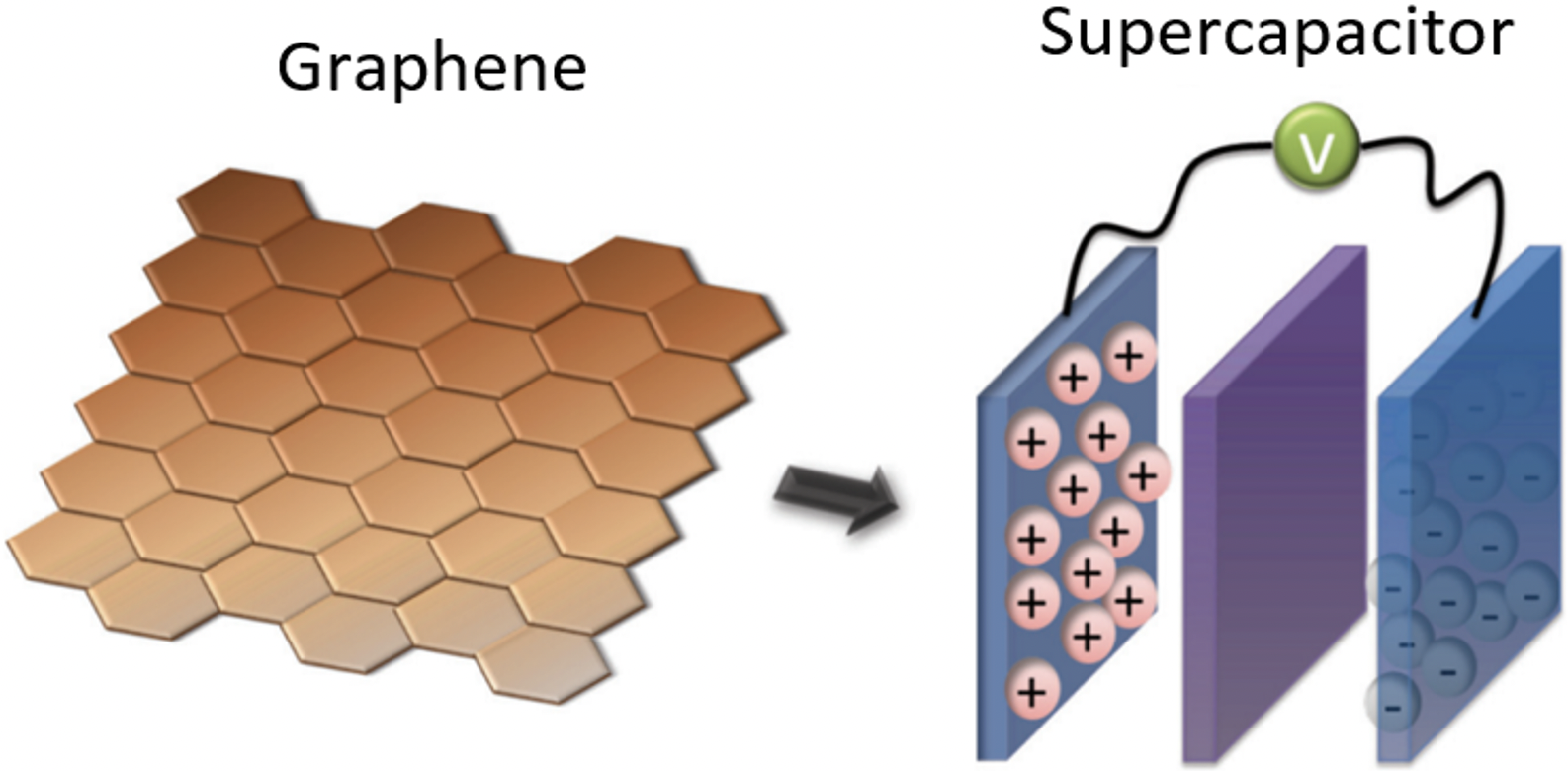}}
\caption{ (a) Single-layer graphene structure. (b) Graphene supercapacitor structure diagram.
(c) Graphene supercapacitor structure diagram\cite{10}.}
\label{fig2}
\end{figure}

There are $\pi$ bonds with weak binding energy in the direction perpendicular to graphene 2D plane(Figure \ref{fig2a}), and electrons can move freely in graphene, and its semi-filled structure makes the electron conduction rate up to 8$\times$10$^5$m/s\cite{5}, graphene has excellent conductivity, these advantages are widely used in supercapacitors and other aspects. Supercapacitors are called electrochemical capacitors, which have the characteristics of high power density, long cycle life, fast charge and discharge, and no pollution to the environment.

The energy storage mechanism of supercapacitors is divided into electric double-layer capacitors and pseudocapacitive capacitors. Electric double-layer capacitors store energy through the double layer formed between the electrode and the electrolyte. The electrode materials are mainly activated carbon, carbon fiber, carbon nanotubes, graphene and other carbon materials; In pseudocapacitive capacitors, if the electroactive material is deposited on the surface or the 2D or quasi-2D space of the electrode material, the Faraday oxidation-reduction reaction occurs, and the capacitance related to the electrode charging potential is generated. The electrode material is mainly metal oxide and Conductive high molecular polymer\cite{9}.

The main disadvantage of supercapacitors is low energy density. The energy density of supercapacitors is 2$\sim$10W·h/kg, which is lower than that of lead-acid batteries (20$\sim$40W·h/kg), nickel-hydrogen batteries (40$\sim$100W·h/kg) and commercial lithium-ion batteries (100$\sim$ 200W·h/kg). Graphene materials with unique ultra-thin 2D structure, excellent conductivity (5000 S/cm), high specific surface area (2620 m$^2$/g), high theoretical specific capacity (550F /g), high area specific capacity (21 F/cm$^2$) and good mechanical properties have been proved to be an ideal material for supercapacitor electrodes. Applied graphene electrode materials to supercapacitors can significantly increase their energy density by more than tens of times and greatly improve their power density\cite{11,12}(Figure \ref{fig2b}-\ref{fig2c}). 
\subsubsection{Doped graphene}
\begin{figure}[h!]
\centering
\subfigure[]{\label{fig3a}
\includegraphics[width=0.44\linewidth]{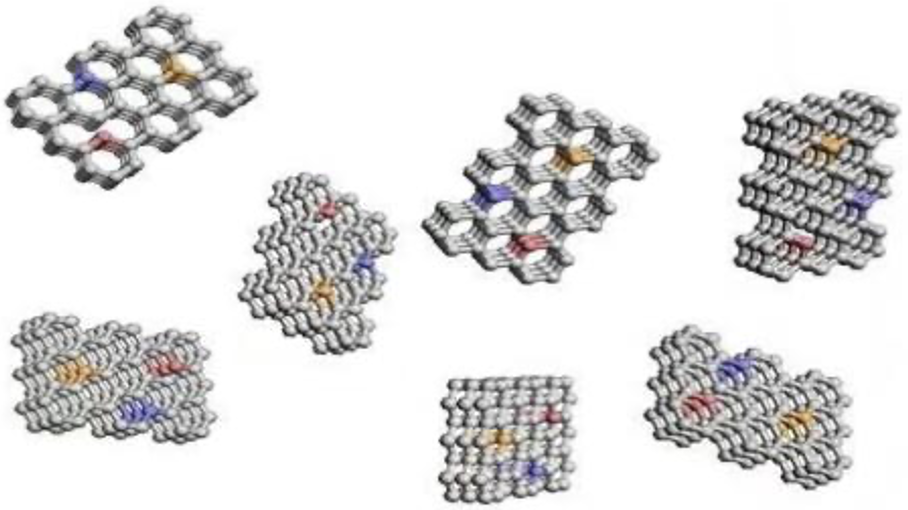}}
\hspace{0.01\linewidth}
\subfigure[]{\label{fig3b}
\includegraphics[width=0.52\linewidth]{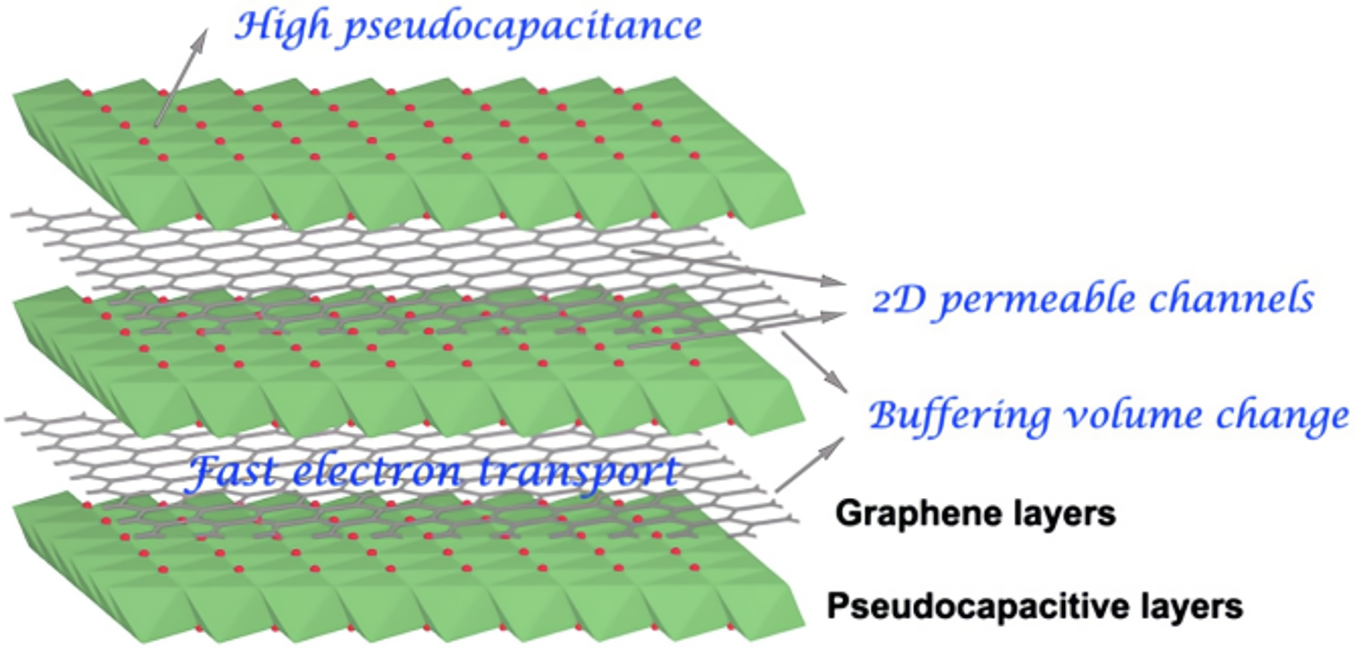}}
   \caption{(a) Graphene stacking and agglomeration\cite{14}. (b) 2D layered-Ni(OH)$_2$ pseudocapacitive material and 2D graphene layer-by-layer composite structure\cite{16}.}
   \label{fig3}
\end{figure}

Doping heterogeneous atoms in the graphene character can significantly improve its electrochemical performance. The introduction of heterogeneous atoms can change the intrinsic physical and chemical properties of graphene, including basic electronic properties, mechanical properties, and hydrophilic and lipophilic properties. Common doping heteroatoms include nitrogen atoms (N), boron atoms (B), sulfur atoms (S), and phosphorus atoms (P). Among them, N atoms are the most widely studied. According to the different positions of N atom doping, graphitized nitrogen, pyrrole nitrogen and pyridine nitrogen can be obtained. The latter two can significantly improve the electrochemical performance of graphene. The specific capacity of graphene electrode materials doped with nitrogen atoms is generally 200-400F/g, which is nearly 4 times higher than that of undoped graphene\cite{13}. In addition to single element doping, two (such as B and N; S and N) or more than two elements can also be doped together to enhance the electrochemical performance of graphene. However, doped graphene cannot avoid the stacking and agglomeration between graphene, and other structural design and assembly methods need to be combined to avoid the problem of graphene restacking and agglomeration\cite{14,15}(Figure \ref{fig3a}).
\subsubsection{New Pseudocapacitive Electrode Material}
Graphene/metal oxide and graphene/conductive polymer are the two types of graphene composite electrode materials that have been studied most deeply. Common metal oxides include manganese oxide (MnO2) and ruthenium oxide (RuO), and conductive polymers include polyaniline and polypyrrole, etc. Metal oxides and conductive polymers can be used as pseudocapacitive electrode materials to undergo rapid and reversible redox reactions on the surface, thereby delivering high specific capacity. However, these materials themselves have disadvantages such as low conductivity and poor cycle performance, which greatly limits their practical application in supercapacitors. In order to improve this situation, graphene with high specific surface area, high conductivity and room temperature inertness is usually used to composite with metal oxides and conductive polymers to form a new type of pseudocapacitive electrode material. This type of graphene composite material combines the advantages of graphene and metal oxides or conductive polymers, and a significant synergistic effect can be produced between the two\cite{14}(Figure \ref{fig3b}).

First, the pseudocapacitive material is loaded on the graphene surface to prevent re-stacking between graphene sheets, which not only facilitates ion transport, but also increases the active specific surface area available for graphene, thereby improving charge storage. Secondly, pseudocapacitive materials can usually be uniformly bonded to conductive graphene surfaces in special nanostructures or particles, not only greatly promoting reversible redox reactions on the surface of pseudocapacitive materials, but also accelerating the transmission of electrons, increasing the specific capacity of pseudocapacitive materials. Furthermore, the pseudocapacitive nanomaterials are anchored on the surface of graphene, which can effectively prevent the particles from gradually agglomerating and growing up and the electrodes powdered or destroyed during the repeated Faraday reaction process, thereby improving the cycle stability of the material. Therefore, the synergistic effect of the composite material can not only increase the conductivity of the oxide or polymer material, the specific capacity of the pseudocapacitive and the electric double layer of graphene, but also greatly improve the cycle stability of the pseudocapacitive electrode material\cite{14}. 

It should be pointed out that the pseudocapacitive material and the graphene composite material undergo different preparation methods, and the mass specific capacity is very different. For example, the mass specific capacity of the polyaniline/graphene composite is 500-1200F/g, and the polypyrrole/graphene composite is 300-600F/g. The mass specific capacity of composite materials tends to increase as the content of pseudocapacitive materials increases, but its conductivity decreases. Compared with pure pseudocapacitive materials, the mass specific capacity of graphene composite materials may be slightly reduced, but its cycle performance and power density will be significantly improved\cite{17}.
\subsubsection{Flexible Super Capacitor}
\begin{figure}[h!]
    \centering
    \includegraphics[width=0.5\textwidth]{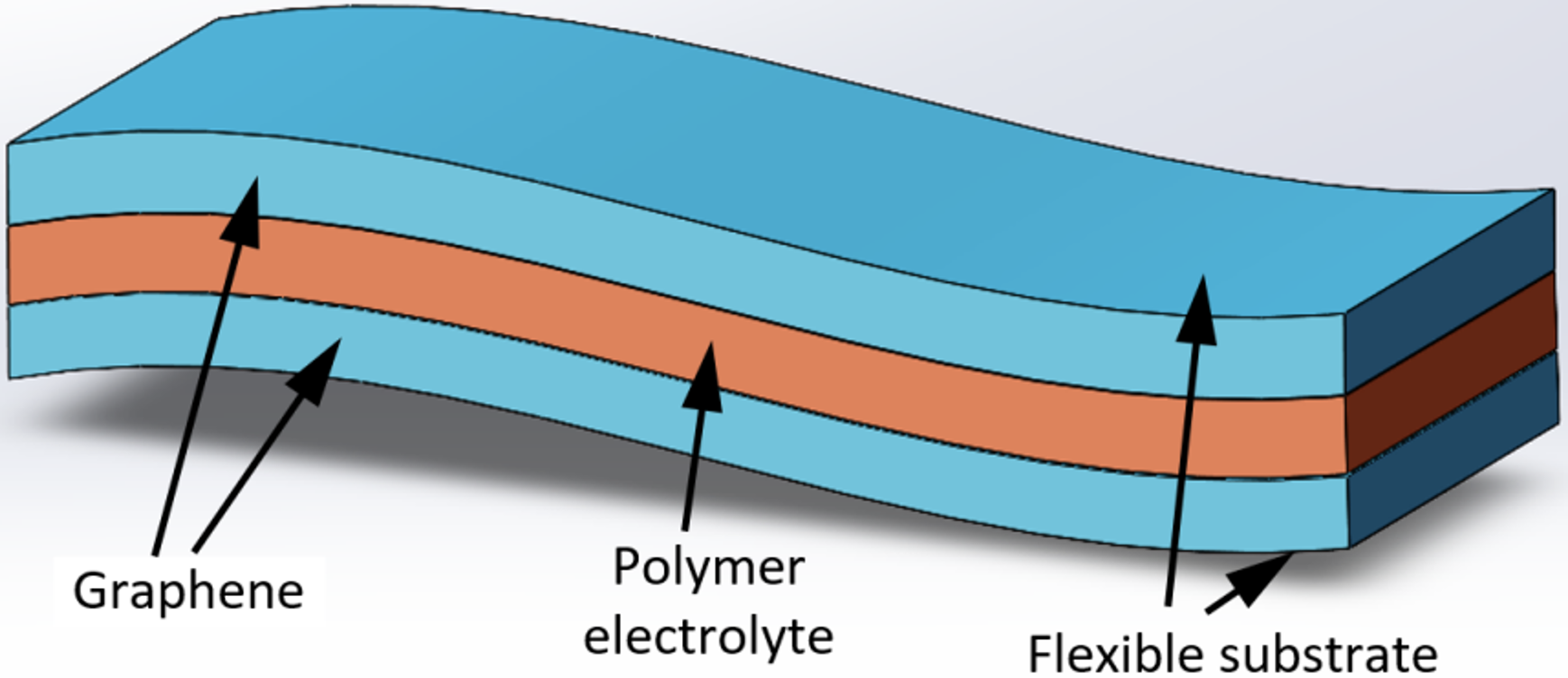}
    \caption{Graphene flexible supercapacitor}
    \label{fig5}
\end{figure}

Flexible supercapacitors are a very promising energy storage device. The key point of its development is to find electrode materials with good flexibility, high conductivity and excellent electrochemical performance. Graphene, especially graphene film and fiber materials are ideal raw materials for preparing flexible electrode materials. Based on graphene materials, macroscopic bulk electrode materials constructed through structural design and assembly, such as one-dimensional graphene fibers, 2D graphene films, and three-dimensional graphene networks, endow the new type of graphene flexible electrodes with unique properties. It has the common characteristics of high specific surface area, developed pore structure, high conductivity, high breaking strength, and no need for additives and conductive agents. Importantly, these graphene flexible electrodes can be used as flexible support substrates and electrode conductive network skeletons, as well as high-performance energy storage electrode active materials, and can be widely used in flexible, bendable, and stretchable supercapacitors\cite{18,19}(Figure \ref{fig5}).

The research on flexible energy storage devices is one of the important directions for the development of energy devices in the future. The research on new types of flexible and even planar energy storage devices conforms to the urgent need for the development of new energy sources in the world today. Although in terms of capacitor performance, supercapacitor devices made of 2D materials as electrode materials have no absolute advantage over three-dimensional supercapacitors. However, when supercapacitors require high flexibility, flatness, and lightness, flexible supercapacitors or planar supercapacitors using 2D materials (such as graphene) as electrode materials have more advantages than other non-2D materials. It is manifested as a higher number of cycles, a higher degree of flexibility, and the performance does not change significantly after several thousand times of folding\cite{20}. This design brings many advantages and lays the foundation for the development of large-scale flexible electronic devices in the future.
\subsubsection{Hybrid super capacitor}
Hybrid supercapacitors generally refer to devices composed of different types of positive and anode electrode materials: one electrode is a battery material containing pseudocapacitors, and the other electrode is a double-layer capacitor material. Hybrid supercapacitors combine the fast charge and discharge characteristics of electric double layer materials and the high energy density of pseudocapacitors. They can have high power density and high energy density at the same time, making up for the vacancy that cannot be achieved in both batteries and supercapacitors. Graphene-based hybrid supercapacitors can obtain high energy density and power density at the same time, combined with the advantages of double layer capacitor and Faraday quasi capacitor, can better meet the overall requirements of the energy density and power density of the power system, suitable for short time large current discharge, can be used as the electric vehicle start and braking power supply\cite{21}, and can also play a great role in electric aircraft.
\subsection{Solar cells and graphene}
\begin{figure}[h!]
\centering
\subfigure[]{\label{fig6a}
\includegraphics[width=0.45\linewidth]{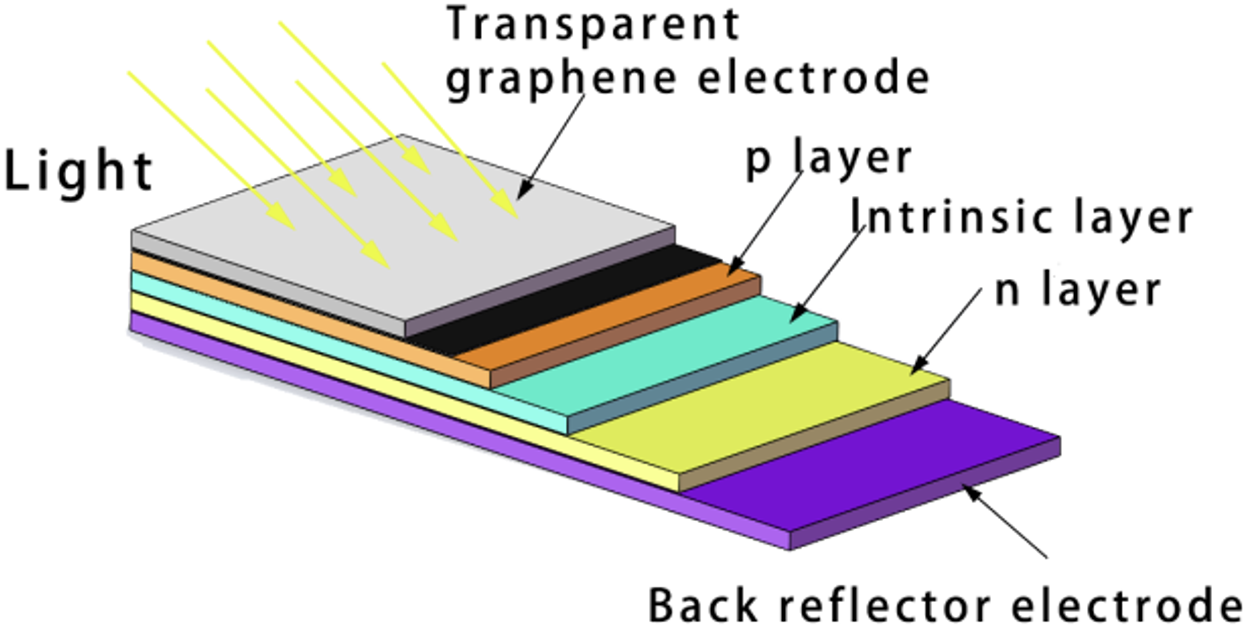}}
\hspace{0.01\linewidth}
\subfigure[]{\label{fig6b}
\includegraphics[width=0.45\linewidth]{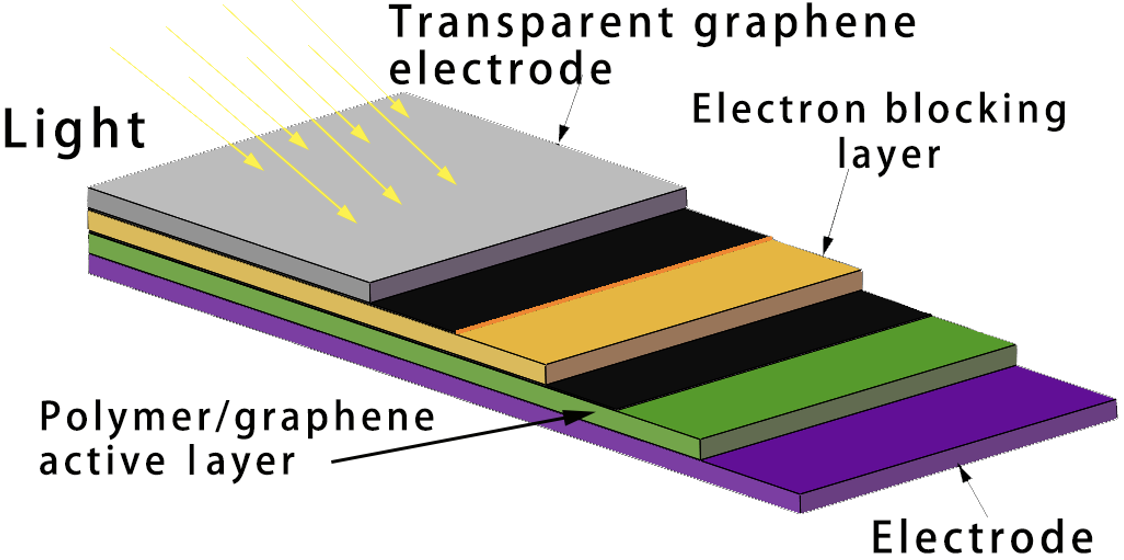}}
\subfigure[]{\label{fig6c}
\includegraphics[width=0.6\linewidth]{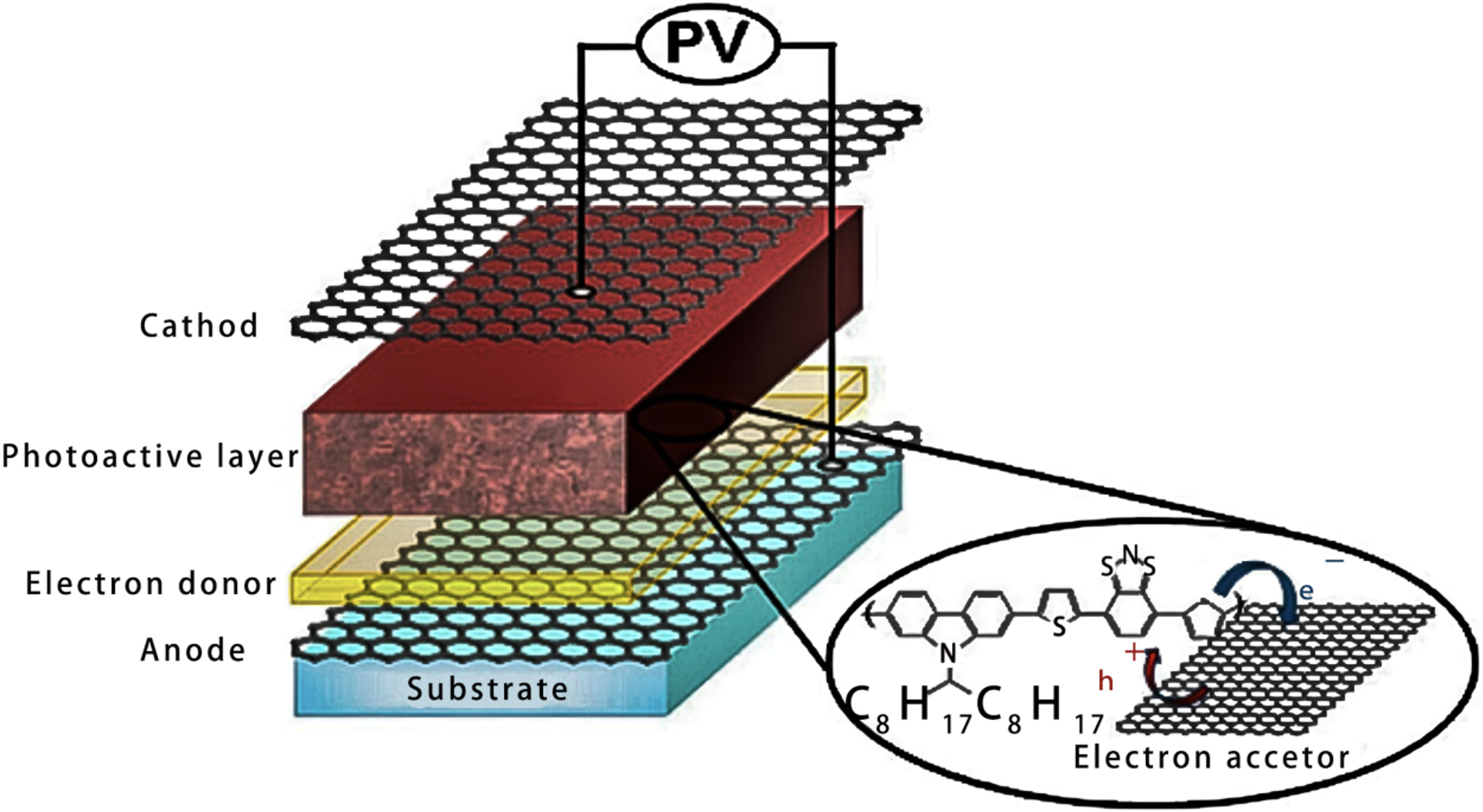}}
   \caption{(a) Schematic diagram of graphene inorganic solar cells. (b) Schematic diagram of graphene organic solar cells. (c) Properly processed graphene can be used as different components in solar cells, including cathode, anode, and photosensitive layer\cite{22}.}
   \label{fig6}
\end{figure}

For decades, with the emergence of new equipment and technologies, storage and effective use of solar energy has been a new way to encourage the exploration of clean energy production. The sun is rich, safe, and can be directly converted into cheap and clean electric energy without pollution and environmental problems. Photovoltaic materials and devices promote the conversion of sunlight to generate electricity through the photovoltaic effect. Due to its unique optical and electrical properties, graphene is a very important material for industrial applications and basic research. Graphene has the characteristics of high light transmittance, high carrier mobility, zero band gap and high mechanical strength. Graphene-based materials have been extensively studied in photovoltaic (PV) technology (Figure \ref{fig6}).
\subsubsection{Dye-sensitized solar cell}
\begin{figure}[h!]
\centering
\subfigure[]{\label{fig7a}
\includegraphics[width=0.45\linewidth]{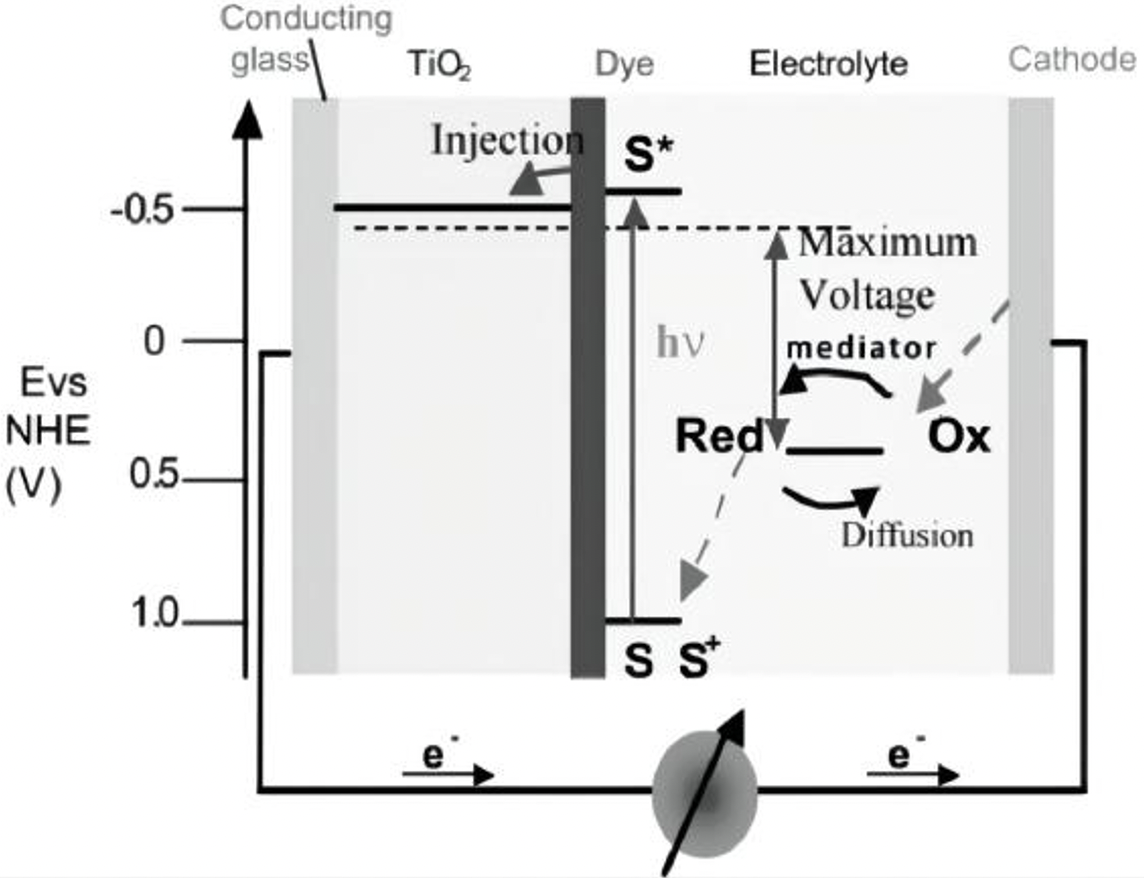}}
\hspace{0.01\linewidth}
\subfigure[]{\label{fig7b}
\includegraphics[width=0.45\linewidth]{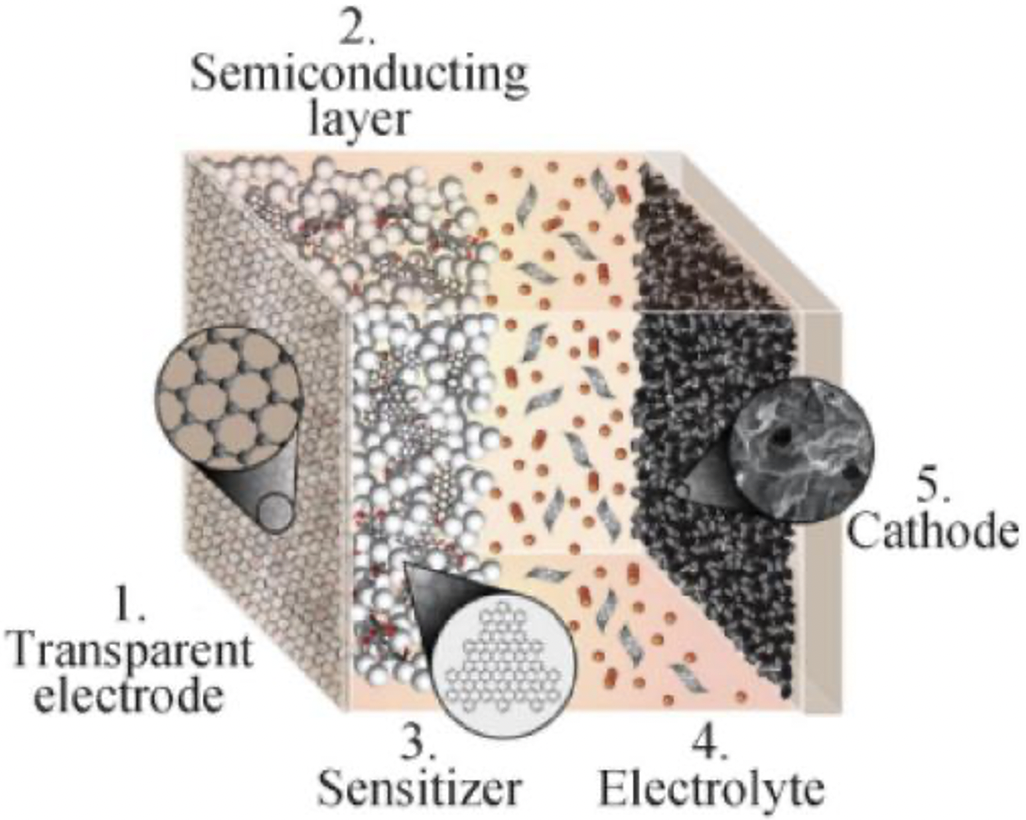}}
   \caption{(a) Working principle diagram of DSSCs. (b) Schematic diagram of the structure of adding graphene at different positions of DSSCs.}
   \label{fig7}
\end{figure}

Dye-sensitized solar cell (DSSCs), as a promising low-cost solar cell, has attracted the interest of academia and industry due to its relatively high photoelectric conversion efficiency, low production cost and environmental friendliness\cite{23,24,25,26}. Its working principle is shown in Figure \ref{fig7a}. The 5 parts of DSSCs are shown in Figure \ref{fig7b}.

As a rising star in the field of materials science, graphene has recently been used as one of the key components of DSSCs that exhibits high-performance characteristics. Considering the photoanode, it is believed that the appropriate addition of graphene in some cases may increase the photocurrent in some cases. However, it is currently unclear whether the selective and careful doping of graphene photoanodes will increase the power conversion efficiency (PCE) of the device. As a sensitizer, graphene shows advantages in multi-carrier generation and thermal injection, and provides a possible method to overcome the inherent limitations of current device structures. The proper concentration of various graphene in the electrolyte (trace additives) can improve the performance of DSSCs, but due to its catalytic activity and high light absorption, high concentration of graphene will reduce the performance of the device. In terms of anode materials, the excellent oxidation-reduction reaction reduction catalytic activity makes graphene and its composite materials a strong candidate for replacing Pt and FTO (SnO2: F) with DSSCs cathodes. According to the development trend, graphene materials will have extensive and important future applications in DSSCs. It should be noted that the application of graphene materials in DSSCs is still in its infancy, and most of the mechanisms discussed are based on various untested assumptions. Before realizing the practical application of graphene materials in DSSCs, various graphene materials with different properties should be extensively studied as different components of DSSCs\cite{27}.
\subsubsection{Perovskite and organic solar cells}
\begin{figure}[h!]
\centering
\subfigure[]{\label{fig8a}
\includegraphics[width=0.4\linewidth]{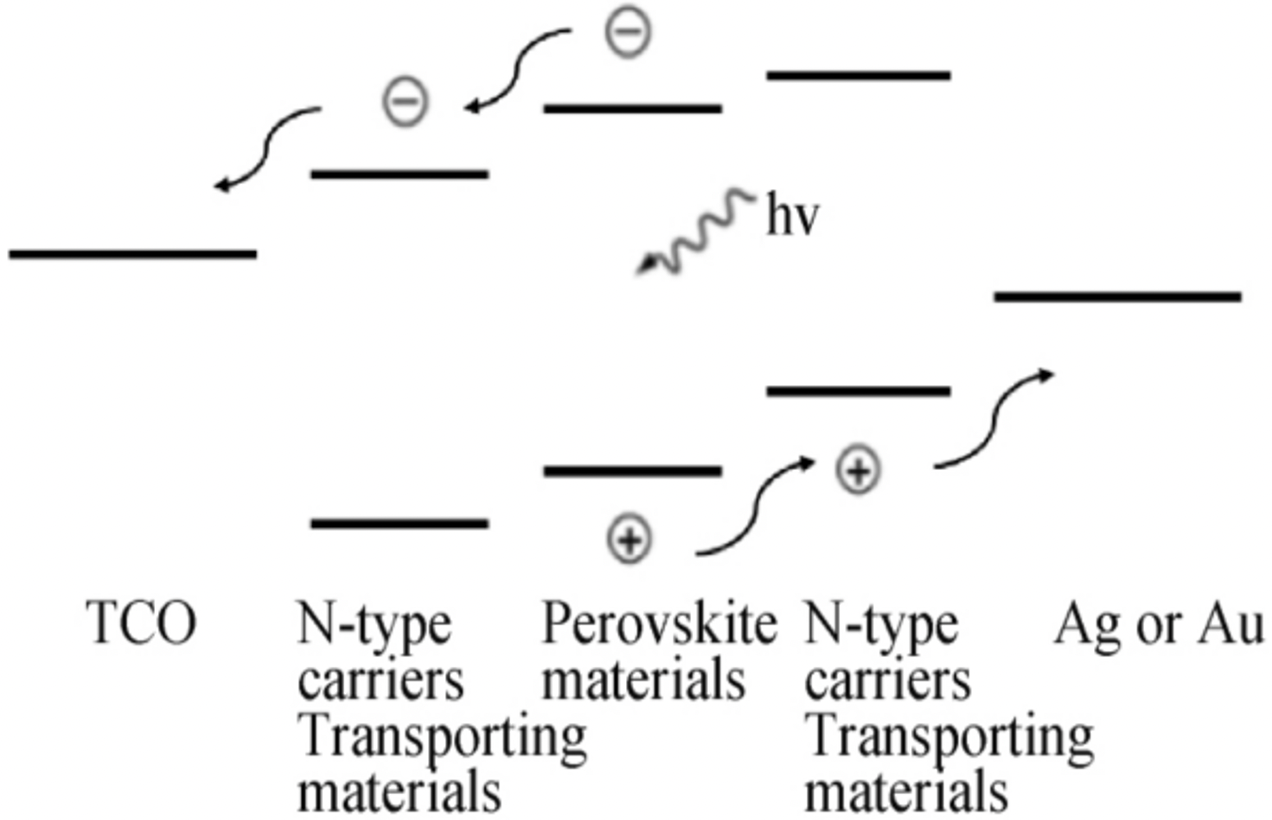}}
\hspace{0.01\linewidth}
\subfigure[]{\label{fig8b}
\includegraphics[width=0.4\linewidth]{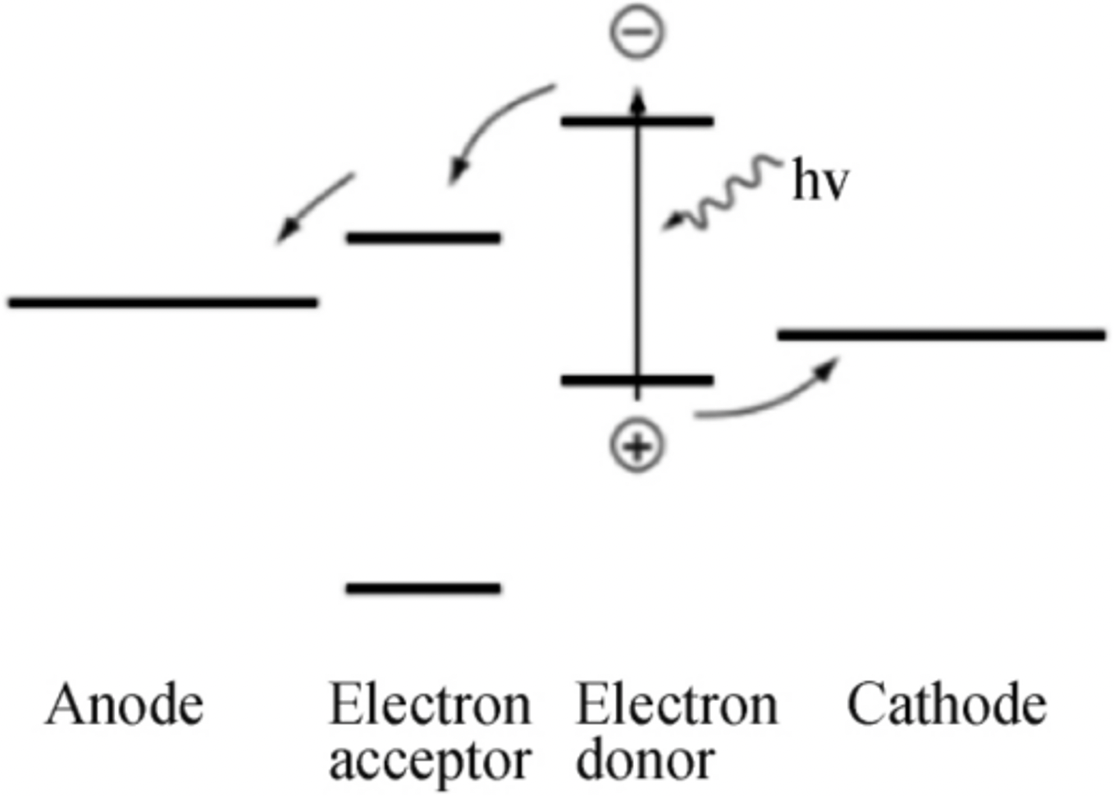}}

\subfigure[]{\label{fig8c}
\includegraphics[width=0.35\linewidth]{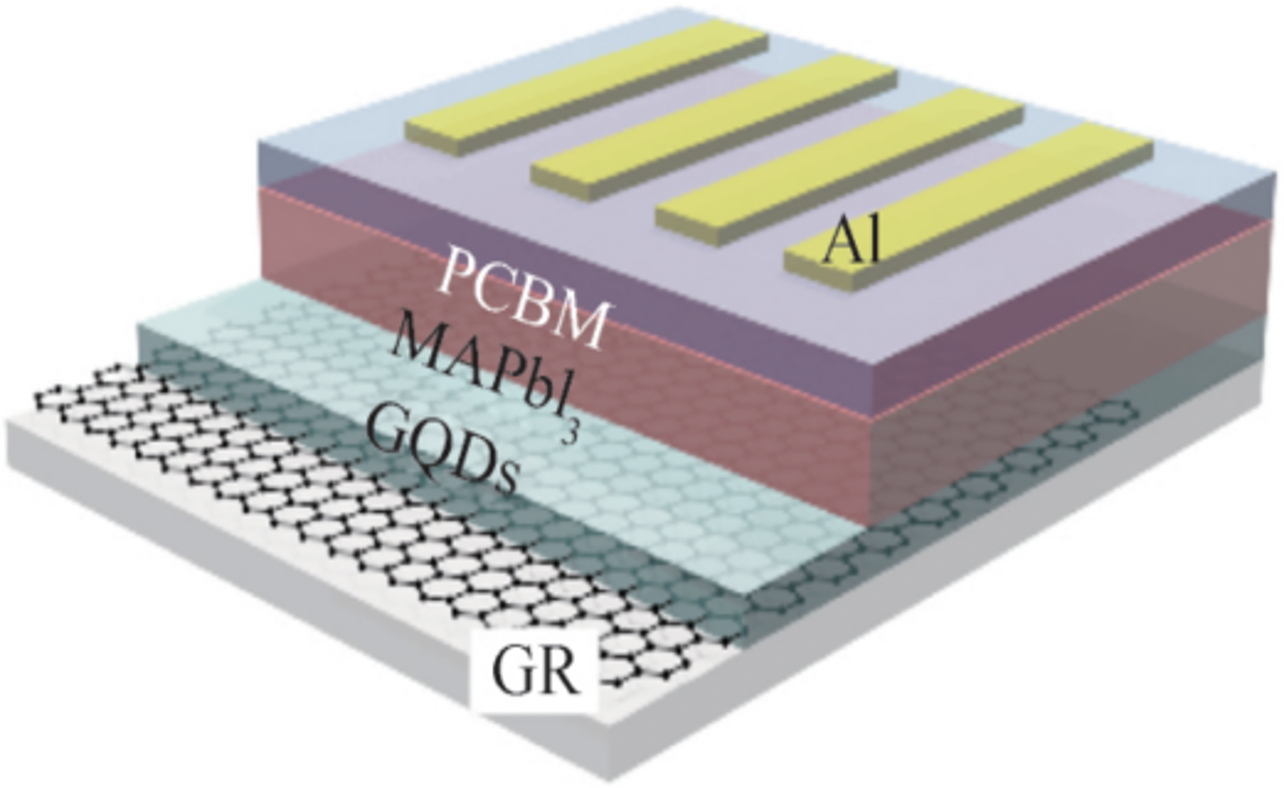}}
\hspace{0.01\linewidth}
\subfigure[]{\label{fig8d}
\includegraphics[width=0.35\linewidth]{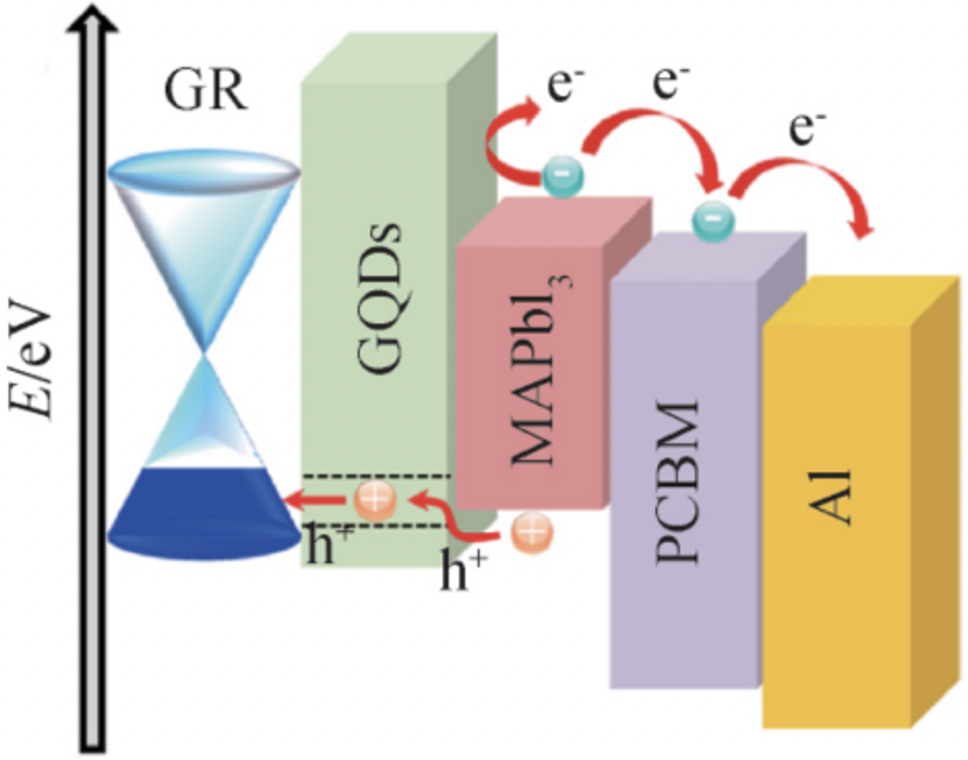}}
   \caption{(a) Working principle diagram of perovskite solar cell. (b) Working principle diagram of organic solar cell\cite{33}. (c) Schematic diagram of graphene as a transparent conductive electrode\cite{37}.
(d) Band structure diagram of graphene-containing battery with p-i-n structure\cite{37}.}
\label{fig8}
\end{figure}

Both organic solar cells and perovskite cells have material structure diversity and low cost. They have developed rapidly in recent years, and the photoelectric conversion efficiency has been continuously improved\cite{28,29,30,31,32}. When it is exposed to sunlight, light is absorbed by the perovskite layer through the transparent conductive film to produce electron-hole pairs, and these uncombined electrons and holes are respectively collected by the electron transport layer and hole transport layer and finally transported to the electrode (Figure \ref{fig8a}). Among them, the reversible recombination of electrons in the electron transport layer and holes in the perovskite layer, and the recombination of electrons in the electron transport layer and holes in the hole transport layer (when the perovskite is not dense), the recombination of the electrons in the perovskite layer and the holes in the hole transport layer, etc., will cause energy loss. To improve the overall performance of the battery, the loss of these carriers should be minimized. The working principle of organic solar cells is similar to the principle of perovskite solar cells\cite{31}, as shown in Figure \ref{fig8b}.

At present, the transparent conductive film (TCE) of solar cells is mainly made of indium tin oxide (ITO). However, due to the limited supply of indium on the earth, issues such as pH sensitivity\cite{34,35} and mechanical fragility\cite{36} limit the application of ITO in photovoltaic devices. Therefore, it is necessary to develop a new type of transparent conductive film, and the conductive film material needs to have the following characteristics: low cost, mechanically strong, transparent, high conductivity, and suitable work function. Graphene materials have been successfully used as transparent electrodes in perovskite solar cells\cite{37,38,39,40,41} (Figure \ref{fig8c}-\ref{fig8d}).

With the continuous progress and update of preparation technology, researchers have conducted a lot of research on the application of graphene and its derivatives in solar cells. Graphene has the unique characteristics of high carrier mobility, low resistivity and transmittance, and 2D network, so it can be used as additives in electrodes, functional layers, absorber layers or semiconductor layers. Especially when used as electrodes in flexible devices, its mechanical stability and flexibility can solve the problems of rigidity and brittleness of traditional TCO. Although graphene-based photovoltaic devices have made significant progress, they still face many challenges in the future. It is necessary to develop a cost-effective technology for mass production of large-scale continuous graphene films, and also consider the influence of the number of graphene layers, doping, and functionalization on the overall performance of the device. In short, graphene has broad application prospects in solar cells, but more basic research and development are still needed to support the large-scale and commercialization of the solar photovoltaic industry\cite{42}.
\section{Graphene sensor applications in the aviation field}
A sensor is a detection device that can perceive the measured information, and can convert the sensed information into electrical signals or other required forms of information output according to a certain rule, so as to meet the requirements of transmission, processing, storage, display, recording and control, etc. requirements. The characteristics of sensors include: miniaturization, digitization, intelligence, multi-function, systemization, and networking. There are about 2,000 to 4,000 sensors on the plane. The main sensors are: airspeed tube, static pressure hole, temperature, stall sensor, gyroscope, angle of attack sensor, ground sensor, humidity sensor, acceleration sensor, etc. They play a huge role related to flight safety. Graphene has its own characteristics such as high conductivity, high toughness, high strength, large specific surface area and other outstanding properties, which can play a significant role in the field of sensors. For example, gas sensor: oxygen monitoring; strain sensor: aircraft load spectrum test; electrochemical sensor: aircraft corrosion detection; magnetic field sensor: aircraft high-precision navigation and positioning; photoelectric sensor: fuel control system for fuel control.
\subsection{Electronic and mechanical properties of graphene}
\begin{figure}[h!]
\centering
\subfigure[]{\label{fig9a}
\includegraphics[width=0.25\linewidth]{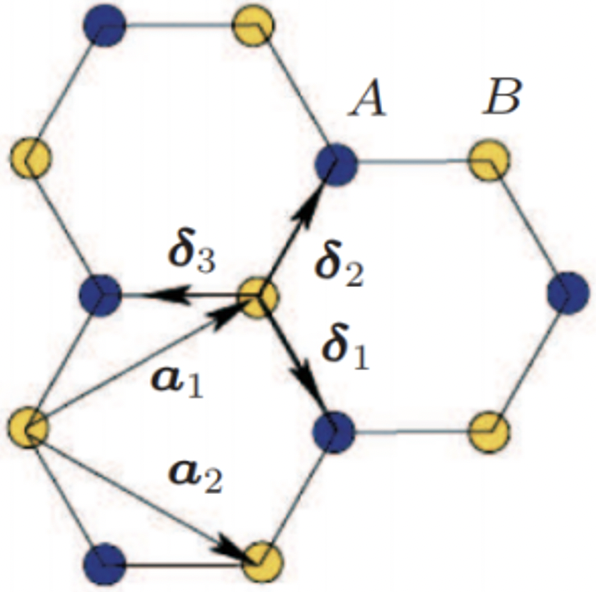}}
\hspace{0.01\linewidth}
\subfigure[]{\label{fig9b}
\includegraphics[width=0.4\linewidth]{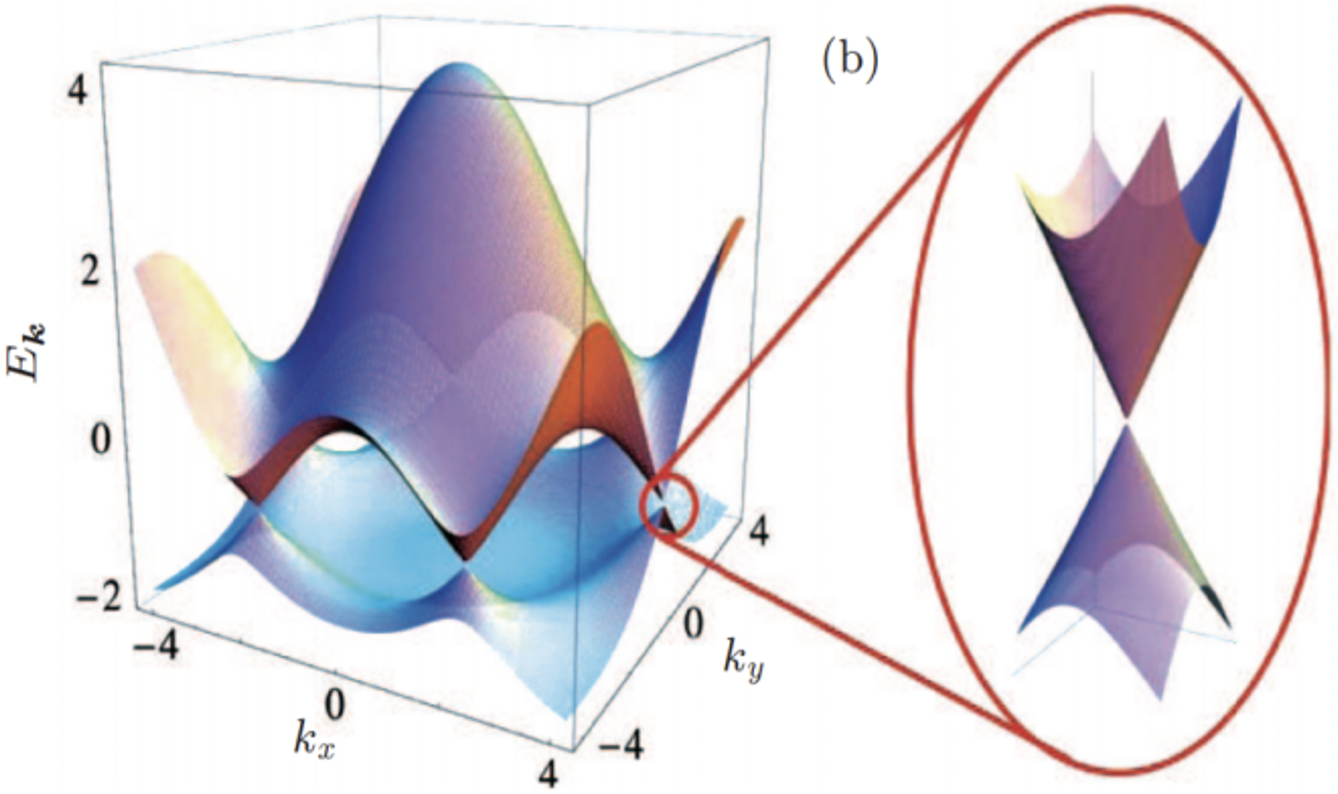}}
   \caption{Graphene band structure. (a) Graphene lattice structure. Each primitive cell contains two symmetric and inequality carbon atoms A and B, and 1n2 is the lattice vector. (b) The graphene electronic band structure obtained in the tight-binding model; the valence band and the conduction band are connected in the Brillouin zone\cite{48}.}
   \label{fig9}
\end{figure}

Graphene is densely packed with sp2 carbon atoms and can be rolled up to form zero-dimensional fullerenes and one-dimensional carbon nanotubes. Because of this structure, the carrier dynamics is strictly limited in the 2D layer. The honeycomb lattice has two equivalent lattice points, A and B, as shown in Figure \ref{fig9a}, leading to special electronic jumps. For single-layer graphene, the electronic band structure according to the tight-binding model is shown in Figure \ref{fig9b}. The valence and conduction bands of graphene are cone-shaped valleys that touch at the Dirac points K and K0 in the Brillouin zone\cite{43}. Near the Dirac point, the carrier has a linear dispersion relationship $E =\hbar v_Fk$, so it can be called a massless Dirac fermion. The speed of electrons in graphene is about 106 m/s, which is about 1/300 of the speed of light. Double-layer graphene is also a zero-band gap semiconductor, but its electronic dispersion is not linear near the Dirac point. For graphene with more than three layers, the energy band structure becomes more complicated, and the valence band and conduction band begin to overlap. Due to its special electronic energy band structure, graphene contains rich and novel physical phenomena and properties, such as ultra-high mobility\cite{44}, ballistic transmission, anomalous sub-Hall effect\cite{45}, non-zero minimum quantum conductivity\cite{46}, Anderson weak local variation\cite{47}, and Klein tunnel\cite{48}.

Scientists' growing interest in graphene is not only driven by its unusual physical properties, but also by its potential to develop various types of sensors. Its mechanical stiffness is about 1TPa, and its inherent breaking strength at 25\% strain is 130GPa, which is comparable to the significant planar value of graphite and other materials with high mechanical strength\cite{48}.
\subsection{Graphene gas sensor}
A gas sensor is an electronic device that can qualitatively or quantitatively detect a specific gas. It has been widely used in many fields such as indoor and outdoor gas monitoring, industrial control, agricultural production, medical diagnosis, military and public security\cite{49,50,51,52}. Generally, traditional gas sensor materials are composed of semiconductor metal oxides, conductive polymers and carbon nanotubes\cite{53,54}. Among them, metal oxides have become the most successful and fully commercialized sensing materials due to their ease of manufacture, high sensitivity and economic cost\cite{55,56}. However, their shortcomings such as high temperature operation, high power consumption and low selectivity are also obvious, making them unsuitable for next-generation wearable sensor applications. Sensors based on conductive polymers can conduct conduction at room temperature without additional power, but will be affected by performance degradation in the air, especially under humidity\cite{57}. The use of carbon nanotubes can greatly reduce the operating temperature of the sensor and bring excellent sensitivity, but its long response and recovery time and complex process hinder its wide application\cite{58}. Therefore, it is very important to develop room temperature working sensor materials with excellent sensing performance for next-generation sensor applications. 

Since the first discovery of graphene, nanomaterials with two-dimensional (2D) structures have aroused widespread research interest worldwide\cite{59,60,61,62,63,64,65,66}. Thanks to its huge surface area to volume ratio, atomic thickness, and excellent electrical conductivity or semiconductor properties, 2D structural materials also show extraordinary potential in the field of gas detection\cite{67,68,69,70}. Their unique 2D structure exposes most of the atoms that can interact with environmental gas molecules and output huge signals. In addition, the ability of 2D nanomaterials to recognize gas analytes at room temperature and their inherent flexibility make them promising candidates for building flexible and wearable gas sensors integrated on low Young's modulus substrates\cite{71}. Nevertheless, before the practical application of 2D nanomaterials in the field of gas detection, many challenges including selectivity, sensitivity, response time, recovery time and stability must be solved\cite{72}. The atomic surface of graphene is chemically inert, resulting in weaker adsorption of gas molecules. Strategies including surface functionalization, foreign atom doping, defect engineering, and ligand conjugation are generally applicable to improve the sensing performance of intrinsic graphene\cite{73,74}. In contrast, the 2D transition metal dichalcogenide (TMDC), an analog of graphene, has a multi-energy band structure, a variety of physical and chemical properties, a layer-related band gap, and excellent catalytic performance. It is more adaptable in the design of actual gas sensing equipment\cite{75,76}. However, due to the strong interaction between molecules and the surface of TMDs, sensors made by 2D TMDCs may be affected by slow response and slow recovery. Incomplete recovery may gradually reduce sensing performance and long-term stability. Recently, research interest has shifted to the development of graphene-like 2D/2D nanocomposites, that is, mixing 2D nanomaterials with other 2D nanomaterials\cite{77,78}.

Table 2: Research on the performance of gas sensors based on 2D/2D nanocomposite materials\cite{79}.
\begin{table}[h]
\tiny
\newcommand{\tabincell}[2]{\begin{tabular}{@{}#1@{}}#2\end{tabular}}
    \centering
    \caption{Research on the performance of gas sensors based on 2D/2D nanocomposite materials\cite{75}.}
    \begin{tabular}{cccccccc}
    \hline
Material&Device type&Synthesis method&Substrate&Analyte&Limit of detection&Working temperature&Response (recovery) time\\\hline
Graphene +MoS$_2$ &Resistive&CVD + mechanical exfoliation&Polyimide&NO$_2$&1.2 ppm&150$^\circ$C&30 min\\
Graphene +MoS$_2$&Resistive&Liquid-phase-co exfoliation&Si/SiO$_2$&Methanol&10ppm&-&210 s (220 s)\\
Graphene +MoS$_2$&Resistive&GA + ATM&Poly-Si&NO$_2$&50 ppb&25$^\circ$C&21.6 s (< 29.4 s)\\
Graphene +MoS$_2$&FET&CVD + mechanical exfoliation&Si/SiO$_2$&NO$_2$&1 ppm&RT&-\\
rGO +MoS$_2$&Resistive&Microwave-assisted exfoliation&PDMS&NH$_2$&0.48 mbar&RT&15 s\\
rGO +MoS$_2$&Resistive&Soft lithographic patterning&PET&NO$_2$&0.15 ppm&90$^\circ$C&-\\
rGO +MoS$_2$&Resistive&Lithography&SiO$_2$/Si&NO$_2$&2 ppm&60$^\circ$C&30 min\\
rGO +MoS$_2$&Resistive&Layer-bylayer self-assembly&SiO$_2$/Si&Formaldehyde&2.5 ppm&RT&73 s\\
rGO +MoS$_2$&Resistive&Self-assembly&PEN&Formaldehyde&2.5 ppm&RT&10 min (13 min)\\
MoS2/WS$_2$&Resistive&Hydrothermal process&-&NO$_2$&10 ppb&RT&1.6 s (27.7 s)\\
rGO +WS$_2$&Resistive&Ball milling and sonication&Si$_3$N$_4$&NO$_2$&1 ppm&RT&22 min (26 min)\\
\tabincell{c}{Defective graphene/\\pristine graphene}&Current&APCVD&Ge&NO$_2$&1 ppm&RT&28 s (238 s)\\
rGO-MoS$_2$-CdS&Resistive&Solvothermal&-&NO$_2$&0.2 ppm&75$^\circ$C&25 s (34 s)\\
BP/h-BN/MoS$_2$&FET&\tabincell{c}{Mechanically exfoliated \\+ e-beam lithography}&SiO$_2$/Si&NO$_2$&3.3 ppb&RT&8 min (8 min)\\\hline
    \end{tabular}
    \label{tab2}
\end{table}

\begin{figure}[htbp!]
\centering
\subfigure[]{\label{fig10a}
\includegraphics[width=0.42\linewidth]{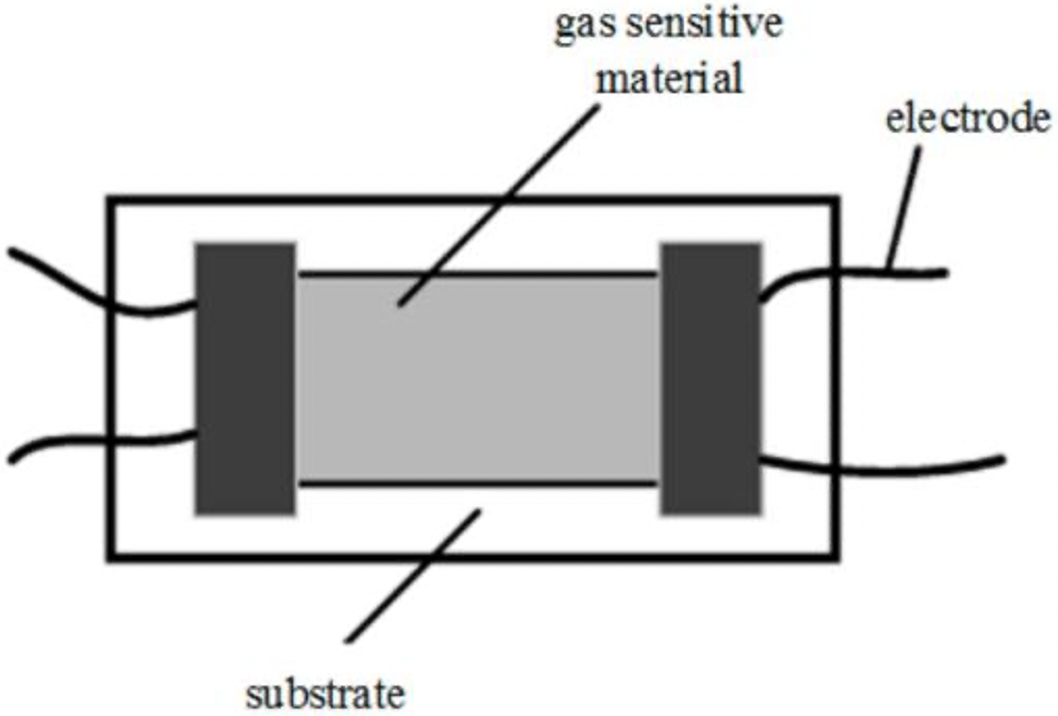}}
\hspace{0.01\linewidth}
\subfigure[]{\label{fig10b}
\includegraphics[width=0.45\linewidth]{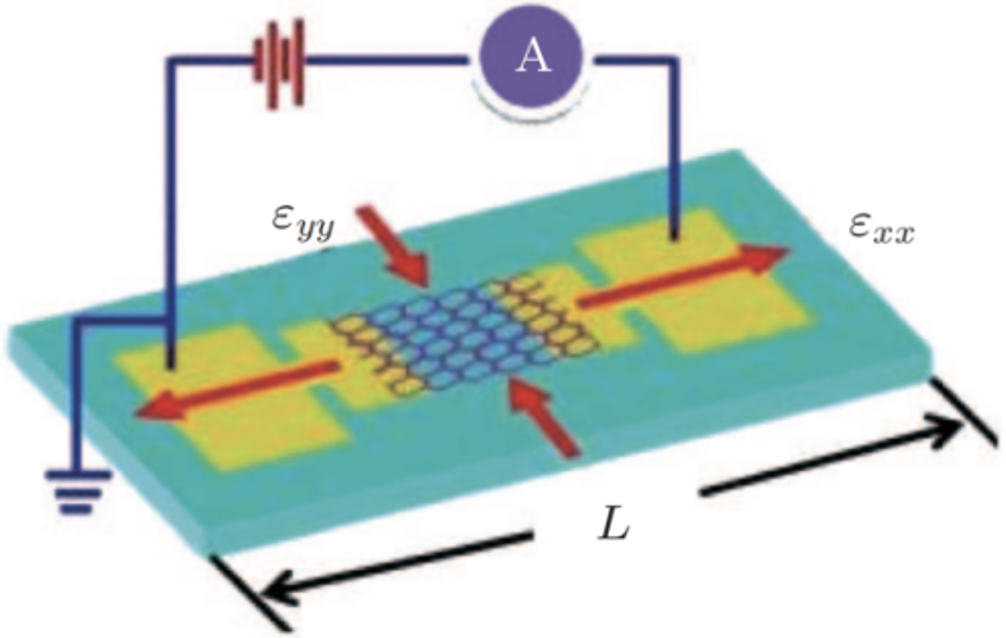}}
   \caption{(a) Resistance gas sensor structure. (b) Schematic diagram of graphene-based strain sensor\cite{87}.}
   \label{fig10}
\end{figure}

So far, most gas sensors are operating in resistance mode. The direct test sensor is used to analyze the gas concentration of the resistance change under the interaction with the detection gas\cite{80}. Its structure is shown in Figure \ref{fig10a}. The substrate is made of insulating materials such as ceramics or silicon dioxide, and graphene materials or various graphene composite materials as gas sensing materials are coated on the surface of the substrate or grown on the surface of the substrate. These electrodes are drawn on both ends of the gas-sensitive material. When the detected gas comes into contact with the gas-sensitive material, gas molecules are adsorbed on the surface of the gas-sensitive material, resulting in a change in resistance. According to the change of resistance, the gas is qualitatively and quantitatively measured\cite{81}. In July 2018, Huang et al. invented a graphene-based atomic oxygen sensor and its preparation method, which proved the feasibility of the graphene-based oxygen sensor and made it possible to apply it in the aerospace field.
\subsection{Graphene-based strain sensor}
Strain sensors monitor fatigue loads experienced by tactical aircraft at various speeds, weights, and mission configurations, and can monitor critical landing gear structures and control surfaces to prevent fatigue damage from high-gravity maneuvers and high-stress landings. With the unremitting pursuit of low cost and miniaturization of devices, traditional silicon semiconductors are facing challenges, and recent research on strain sensors has focused on nanomaterials. Similarly, low-dimensional carbon\cite{82,83,84,85,86} carbon nanotubes (CNTs) and graphene are two examples of the carbon family, which have attracted great attention in recent years due to their outstanding mechanical and electrical properties. Compared with carbon nanotubes with a quasi-one-dimensional (1D) structure, graphene has an ideal two-dimensional (2D) structure, therefore, it has advantages in scalable device manufacturing through a top-down method, and is compatible with existing semiconductor manufacturing technologies. Compared with other material-based strain sensors, this ultra-thin transparent graphene device is more commercial and easily available. The schematic diagram of the structure of the graphene strain sensor is shown in Figure \ref{fig10b}. It includes two electrodes on each side of the graphene\cite{87}. When stress is applied to it, a change in resistance can be observed.

Graphene-based strain sensors have attracted much attention due to their potential applications in miniaturized, high-performance strain sensors, which will break the limitations of traditional silicon-based strain sensors. Even so, there is still a long way to go to realize commercial and sensitive graphene-based strain sensors.
\subsection{Graphene-based fiber grating sensor}
Fiber grating is a micro-optical component that has developed rapidly in recent decades. It has opened up a new application research field for fiber-optic sensing technology. The optical fiber is divided into three parts: core, cladding and coating layer from inside to outside along the radial direction. With a special ultraviolet light irradiation process, the refractive index of the fiber core is permanently perturbed to form a passive optical fiber device, and it can partially or completely reflect the light of a certain wavelength in the incident light, and the wavelength that meets the Bragg condition is reflected by the fiber grating. 

Fiber gratings have a certain sensitivity to changes in the surrounding environment. Zhang et al\cite{88}. discussed the influence of graphene on the optical field transmitted by optical fiber and the response of graphene to gas molecules. It shows that when graphene is combined with fiber grating, it can not only greatly enhance the evanescent field transmitted along the surface of the micro-nano fiber grating, but also change its own dielectric constant through its adsorption of molecules, thereby changing the effective refractive index of the entire hybrid waveguide, causing the corresponding wavelength drift and attenuation, realize the concentration sensing of external micro-molecules\cite{89}. By stretching and compressing the fiber grating, or changing the temperature, the period and effective refractive index of the fiber grating can be changed, so as to achieve the purpose of changing the reflection wavelength of the fiber grating. The reflected wavelength has a linear relationship with the physical quantities of strain, temperature and pressure. According to these characteristics, the fiber grating can be made into a variety of sensors such as strain, temperature, pressure and acceleration.

The aerospace industry is an area where sensors are used intensively, and it has very high requirements for sensor sensitivity, volume and weight. In order to monitor pressure, temperature, vibration, fuel level, landing gear status, wing and rudder position, etc., an aircraft needs to use more than 100 sensors. Therefore, the size, weight, and integration of sensors become important. The fiber grating sensor has only one optical fiber, and the sensitive element (grating) is made in the core. From the advantages of small size and light weight, almost no other sensor can compare with it. 

The application of fiber optic sensors in composite structures has been developed in the aerospace industry, and these industries clearly need these light-weight, rugged and durable sensors. The representative fact is that the optical fiber sensors have been all embedded in the laminated graphite composite board between the layers arranged in the same direction, and all the embedded optical fiber measurement sensors can directly measure the internal stress. The data from the fiber optic sensor is especially decisive under re-entry conditions, and the electrical signal in this case is often blurred due to radio frequency interference. The fiber grating is embedded in a large aircraft, combined with advanced fiber optic communication technology, can form a super distributed intelligent sensor network, which can monitor the internal mechanical performance and external environment of the large aircraft in real time, such as the stress and strain of the large aircraft , Temperature, structural damage and other health conditions.

According to the measurement results, the following functions can be realized:

(1) Automatic alarm and positioning of structural health

The fiber grating structure health monitoring system automatically conducts real-time temperature and strain inspections on the installation area of the fiber grating sensor, detects abnormal fluctuations in the monitored parts of the large aircraft, and detects and alarms in time before possible accidents;

(2) Status query

The monitoring information of each monitoring point is saved to the large-capacity memory of the fiber grating demodulator, the system divides the data into historical information and real-time information according to time, and operation and maintenance personnel can dynamically adjust the real-time status monitoring time interval of monitoring points according to different periods to meet actual requirements. Managers and operators can view the physical parameter change curve of each monitoring point to provide data support for decision-making and maintenance;

(3) Alarm condition setting

The designer can set the alarm trigger conditions of the fiber grating structure health monitoring system to adapt to the differences in monitoring parameters of different parts and different flight conditions;

(4) Line inspection and fault location function

The analysis function of the optical fiber grating demodulator can accurately locate the loss and breakpoint position of the optical fiber transmission line, which is convenient for system debugging, maintenance and line overhaul.

Although fiber grating sensing technology is still in the stage of rapid development, it is foreseeable that with the commercialization and performance of fiber grating sensors, graphene fiber grating sensors will surely show great vitality in the field of sensing. It plays an important and irreplaceable role in the construction of national defense and national economy, especially in the structural health monitoring of large aircraft, which has a very broad application prospect\cite{89}.
\subsection{Graphene-based pH sensor}
Corrosion is a material degradation phenomenon that affects all metals and structures\cite{90}. Structural maintenance and effective management of structural components during their life cycle require timely and accurate corrosion detection\cite{87}. Metal materials in aircraft structures, especially aluminum and steel alloys, are susceptible to time-dependent corrosion, which is usually a slow process of material degradation. Extensive corrosion may occur in aircraft structures, including general corrosion, pitting corrosion, stress corrosion cracking, environmental embrittlement, corrosion fatigue and spalling. The selection of specific metal materials to meet aircraft design requirements is mainly based on their performance attributes, such as weight, stiffness, strength, electrical characteristics, etc., rather than their ability to resist corrosion\cite{91,92}.

By including graphene membranes that are sensitive to specific ions, molecules, and environmental parameters, aircraft corrosion monitoring technology can benefit from the selective sensitivity of functionalized graphene. For example, GO is sensitive to moisture and has been used in various sensors such as capacitive sensors, surface acoustic wave and RFID-based sensors\cite{93,94,95}. According to reports, Reduced graphene oxide (rGO) is a well-balanced temperature sensor with good sensitivity, repeatability and environmental stability\cite{96}. Graphene can also be used as a sensing material in ion selective field effect transistors (ISFET) to measure the content of low-concentration metal ions. Alves et al\cite{97}. A graphene-based metal ion (Na, Co, Al, Cu) sensor is proposed, in which graphene is functionalized by L-phenylalanine, and L-phenylalanine is attracted and combined with metal ions. The sensor shows high sensitivity to a variety of metal ions and displays real-time response\cite{98}. On the other hand, pristine graphene is highly reactive to oxygen-carrying groups such as OH- and has effective quenching ability, making it suitable for chemical and optical pH sensing\cite{98}.

In corrosion monitoring, pH changes are an important parameter indicating corrosion events and their progress. Therefore, monitoring the pH value will help maintain the safety and durability of the aircraft structure. Over the years, several pH sensing methods have been developed, the most common methods are based on paper strips and glass electrodes. However, the accuracy of these methods is limited by their fragility, low sensitivity and unsuitability for miniaturization\cite{94}. Emerging pH sensing technology focuses more on highly sensitive miniature rugged devices that can provide in-situ long-term pH monitoring in harsh environments, such as optical fibers, potential sensors, and ion selective field effect transistors (ISFET)\cite{98}. Currently, ISFET is the most effective alternative to glass probes because of its small footprint, ruggedness and long-term durability. As mentioned earlier, the performance of field-effect transistors largely depends on the sensitivity of the active channel and its responsivity. In this effect, many materials used in pH sensors have been reported in the literature, such as alumina, poly (methacrylic acid) (PMAA), silicate, hydrogel, and fluorescent chips, but due to their low sensitivity and poor degradation, their commercialization is hindered and the difficulty of mass production. Since graphene is successfully separated from highly oriented pyrolytic graphite, compared with glassy carbon and carbon nanotube electrodes, graphene has excellent electrochemical responsiveness and high reaction to oxygen-carrying groups (such as OH-) with high performance and efficient fluorescence quenching ability, it has become a promising candidate for pH sensing applications\cite{98}.
\section{The application of graphene and its composite materials on the exterior of the aircraft body}
\subsection{Lightning protection problem}
\begin{figure}[h!]
\centering
\subfigure[]{\label{fig11a}
\includegraphics[width=0.55\linewidth]{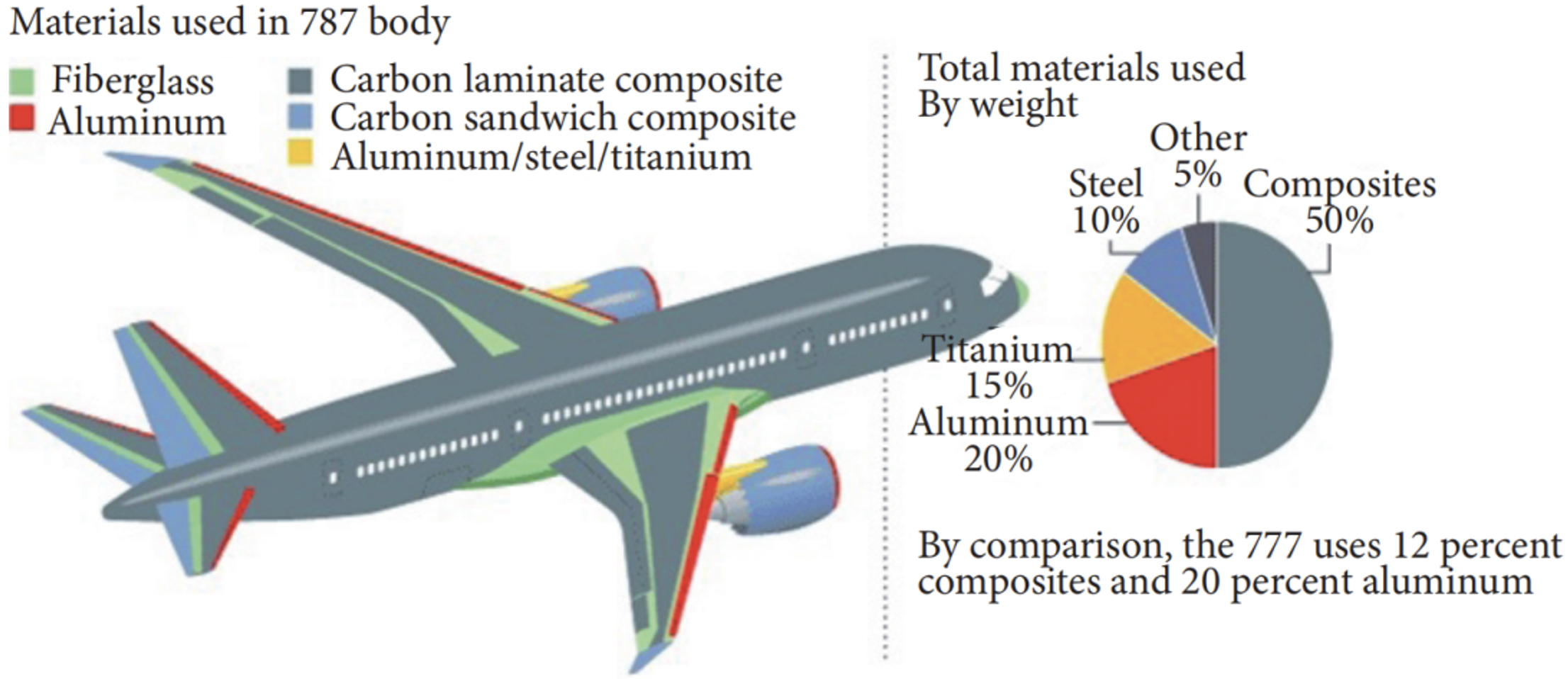}}

\subfigure[]{\label{fig11b}
\includegraphics[width=0.43\linewidth]{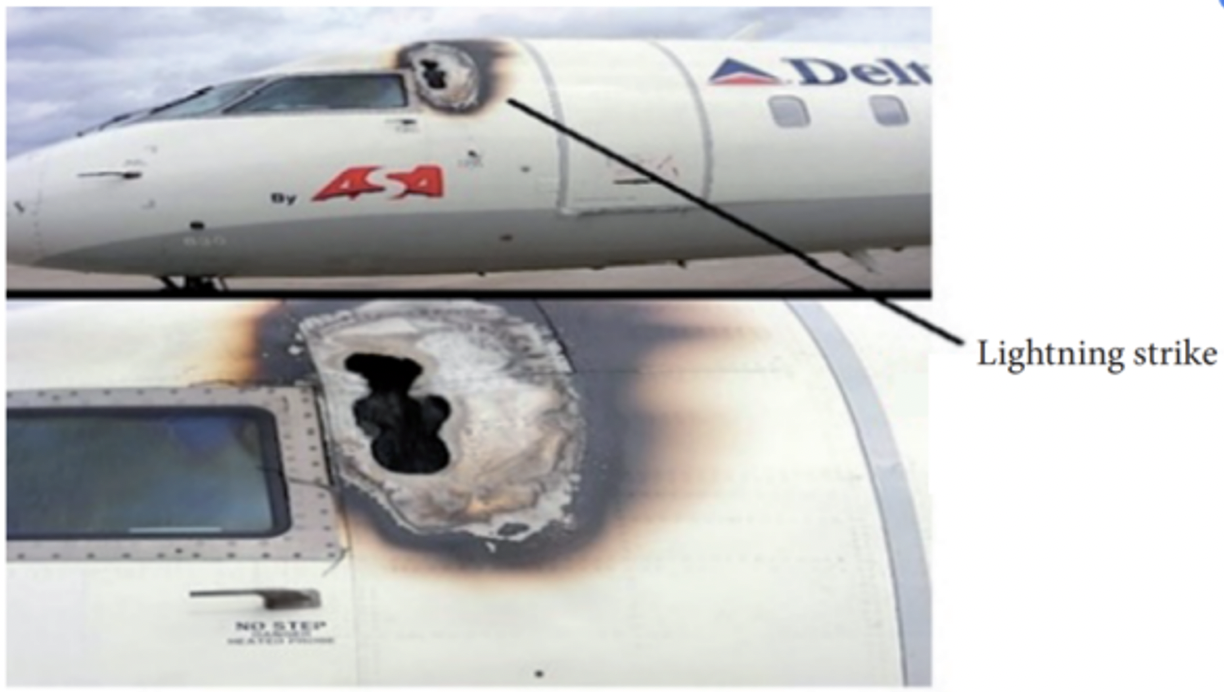}}
\subfigure[]{\label{fig11c}
\includegraphics[width=0.4\linewidth]{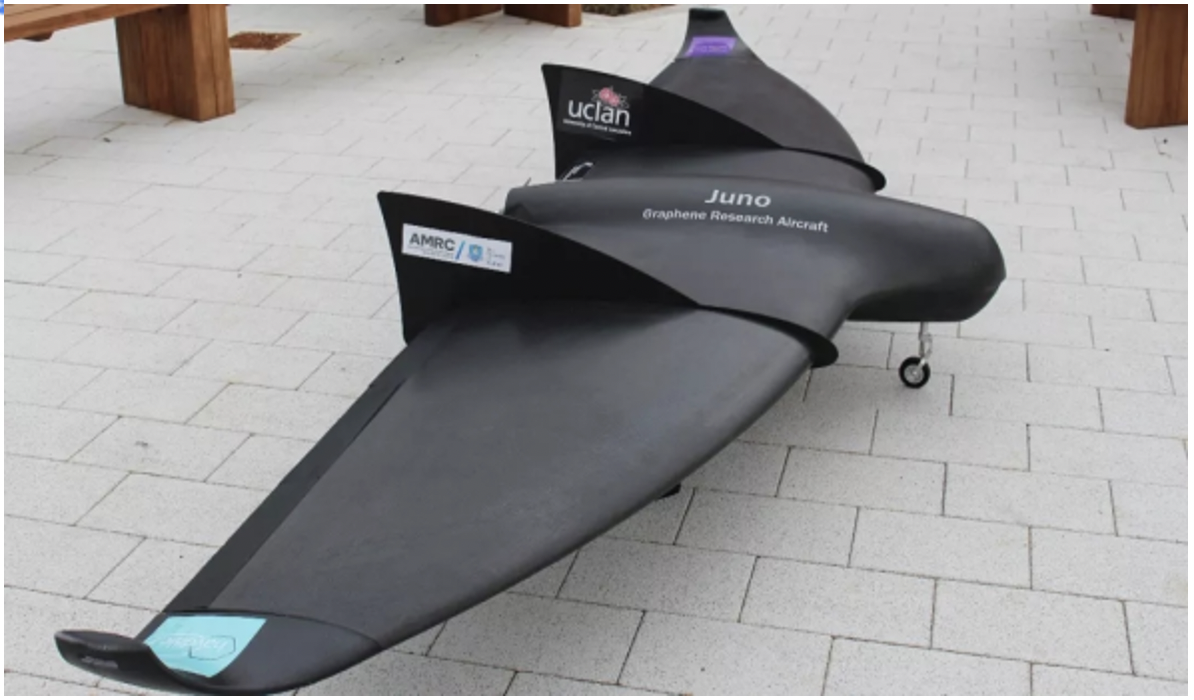}}
   \caption{(a) Materials used in the structure of the Boeing 787 fuselage\cite{99}. (b) Lightning strike effect\cite{100}. (c) The world's first drone with a graphene coating on its outer surface "Juno".}
   \label{fig11}
\end{figure}

In aircraft applications, the weight of the structure is a key parameter that determines its performance. The need for the lowest possible structural weight has led to the development of high-performance composite materials using carbon fiber and epoxy resin. In the past few years, their use has gradually increased, leading to a decline in the use of metal materials. Compared with traditional aircraft materials such as aluminum, titanium, and steel, composite materials are lighter in weight, have higher strength and stiffness values, and are also corrosion resistant\cite{99}. The materials used in the structure of the Boeing 787 fuselage\cite{99} are shown in Figure \ref{fig11a}.

The aircraft exposed some important problems during the flight, these problems will lead to its performance degradation, and even major accidents may occur. One of the problems is that lightning strikes will hit the surface of the aircraft\cite{100}(Figure \ref{fig11b}). Take the fuselage, for example, where high current will be generated. If it is not dissipated in time, it will enter the aircraft and cause a dangerous fire. As the risk of lightning strikes and icing of aircraft in the air increases, resulting in a significant decline in its performance, it is necessary to find a new type of conductive composite material to solve these problems\cite{99}. The composite materials currently used to make aircraft structures do not conduct electricity well; therefore, they are repeatedly damaged by weather conditions. Through the discovery and development of nano-sized reinforcing materials, carbon nanotubes (CNT), nanofibers and graphene are considered to be the key elements of the next generation of reinforced composite materials\cite{101}. By incorporating these nano-reinforced materials into the polymer matrix, not only can the mechanical reinforcement material be reinforced, but other key properties such as electrical conductivity and thermal conductivity can be enhanced without adding additional fillers. Therefore, the use of these conductive composite materials in the aviation industry will effectively solve these problems\cite{99}.

Aircraft composite materials for lightning protection are generally divided into conductive and non-conductive types. Conductive composite materials are used to protect internal flow bands or damage and prevent electric field penetration. On the other hand, these are also non-conductive composite materials, and they cannot conduct lighting current because they are electrical insulators. Therefore, the electric field of a lightning strike can easily penetrate these materials. To prevent this problem, conductive coatings or paints combined with non-conductive composite materials can prevent electric field penetration. In addition, the application of non-conductive composite materials is usually used for secondary structures such as fairings, fins, wings and radomes, or for enclosures that are transparent to radar and radio waves\cite{102}.

In order to reduce the damage caused by lightning strikes to aircraft, researchers have developed various methods of nanomaterials. For example, a conductive carbon nanotube (bucky paper) rubber paper was developed to resist and disperse the effects of lightning; Bucky paper is an effective mechanical, electrical, thermal and other physical properties, and it is a viable engineering material. These materials are used in many applications, including actuators, capacitors, electrodes, field emission devices, radio frequency filters, strain sensing, etc.\cite{103,104}. Similar to foam paper is graphene oxide (GO) paper, which is oriented and assembled by flowing in the water medium of a single graphene oxide sheet. This graphene oxide-based paper material has attracted widespread attention due to its excellent strength, stiffness and high flexibility\cite{109,110,111,112}. This paper-like material is used as sealants, sensors and actuators, radiators, compatible substrates, ion conductors and supercapacitors, flexible substrates with high chemical and thermal stability, electrochemical energy storage devices, structural composite materials and fuel cell\cite{113,114,115,116}. Generally speaking, the energy band gap structure of graphene is zero, which means that graphene has higher conductivity than carbon nanotubes. graphene thin film (GTF) should be added to the composite through a 3D composite mechanism to prevent delamination and debonding between the film and the carbon fiber prepreg, because the electrical properties of GTF before and after being incorporated into the composite are similar\cite{117}. This makes it possible to create a highly conductive graphene thin film (GTF) to reduce the effects of lightning strikes on composite aircraft.
\subsection{Graphene modified fluorocarbon anticorrosive coating protects the surface of aluminum alloy}
In the space shuttle, the most serious corrosion is the structure of the aircraft fuselage. During the take-off and climb phases, due to the temperature difference between the ground and high altitude, a large amount of condensation formed on the surface and interior of the aircraft fuselage. When the relative humidity of the atmosphere in the external environment exceeds 65\%, a water film of about 0.001$\mu$m thick is deposited on the surface of the fuselage and structural parts\cite{118}. The thickness of the water film gradually increases with the increase of the external relative humidity. When the external environmental humidity reaches 100\%, condensation will form on the surface of the fuselage. Corrosion is caused by the contact of saturated chloride ions, oxygen and other corrosive media on the aluminum alloy structural parts of the aircraft. At present, the common types of corrosion of aircraft aluminum alloy structural components include: (1) Pit-like corrosion: In a neutral aqueous solution or humid environment, corrosion pits appear on the surface of aluminum alloy. Their depth and diameter increased rapidly and continued to develop within the aluminum alloy. In the marine environment, chloride ions can accelerate the corrosion of aluminum alloy, and the surface of aluminum alloy is prone to electrical effects; (2) Crevice corrosion: Because the aluminum alloy is in the interfacial gap, the oxidation reaction of the anode solvent and the cathode is reduced, which leads to the accumulation of corrosion products and hinders the transmission of corrosive media; and leading to the difference in corrosive media and concentration inside and outside the crevice, which evolves into "closed battery corrosion";(3) Friction corrosion: chemical and electrochemical reactions occur on the contact surfaces of two aluminum alloys to form corrosion products. The friction between the contact surfaces of the components causes the corrosion products to fall off. As a result, the newly exposed aluminum alloy surface is repeatedly corroded, causing damage to the aluminum alloy structural parts. Organic coating protection is the simplest, most efficient, and most economical anti-corrosion method in the industry\cite{119}.

The development level of heavy-duty coatings is an indicator to measure the advanced level of a country's coatings industry. Aiming at the phenomenon that aluminum alloys are easily corroded, a series of heavy-duty anti-corrosion coatings have been developed at home and abroad. These mainly include: (1) Epoxy heavy-duty anticorrosive coating, a macromolecular chain of epoxy resin, which contains two or more epoxy groups. Thermosetting resins have high viscosity and are extremely hard. Therefore, it is widely used in building materials. However, epoxy resin has extremely high porosity, is easily corroded by oxygen, water, and chloride ions, and has poor acid resistance. Other disadvantages include poor ultraviolet rays and brittle paint film; (2) Polyurethane heavy-duty anticorrosive coating. Due to the reactive cyanate ester groups (-NCO) on the macromolecular chain, this material has extremely high flexibility and high viscosity, and has excellent wear resistance. In the field of heavy anti-corrosion, because of its mechanical properties, it is often used as a top coat to protect the substrate, but its anti-ultraviolet performance is poor, and the anti-corrosion performance is greatly reduced after ultraviolet aging; (3) Fluorocarbon coatings containing a large amount of C, and fluorocarbon resins with a binding energy of up to 485.6 kJ/mol F bond, have excellent chemical resistance and UV resistance. They are widely used in ships, pipelines and aerospace fields, but their adhesion and pigment wettability are poor\cite{120,121}.

In order to find better materials, a lot of research by scientific researchers. For example, Pourhashem et al\cite{122}. used silane coupling agents APTES (KH550) and GPTMS (KH560) to prepare graphene oxide modified epoxy resin coatings to test their protective ability to stainless steel; Ye et al\cite{123}. Covalent grafting is used to prepare functionalized graphene modified epoxy resin coating to explore the corrosion resistance of Q235 steel; Uzoma et al\cite{124}. Use solution polymerization of acrylic monomers to synthesize hydrophobic organics. The siloxane-acrylic resin and fluorosilane modified graphene nanosheets are covalently bonded to prepare a super-hydrophobic ($\ge 152 ^\circ$) anticorrosive coating, which improves the protection of LY12 aluminum alloy. In the above research of graphene modified anti-corrosion coating, graphene oxide (GO) was used to modify the anti-corrosion coating, the protection mechanism of stainless steel was explored, and the reduced graphene oxide modified anti-corrosion coating was used for aviation aluminum material alloy substrate\cite{125,126}.
\subsection{Application of Graphene in the Formulation of Aviation Tire Tread Compound}
Aviation tires\cite{127} are key components that affect the overall performance and safety of aircraft. The tire is equivalent to a cushion air cushion during the take-off and landing of the aircraft. The tire is equivalent to a cushion air cushion during the take-off and landing of the aircraft. The tire needs to be stable to carry the entire load of the aircraft body during taxiing and stop. Especially the tires should be able to effectively resolve the impact energy and absorb the impact energy during the take-off and landing process, and export friction and pressure heat energy, and at the same time should have auxiliary braking force; The working environment of aviation tires requires that they have the characteristics of large load capacity, high inflation pressure, fast heat generation, temperature rise, high impact speed and short allowable sliding distance. 

Therefore, higher standards are set for comprehensive functional indicators such as strength, wear resistance, heat dissipation, grip, air tightness, and static conductivity of aviation tires; With the development of large aircraft, especially military aircraft, more stringent requirements have been placed on the comprehensive functional indicators of aviation tires. However, in the existing technology, aviation tires have problems such as poor wear resistance, short use cycle, impact friction block, poor heat dissipation leading to a decline in tire strength or even flat tire and weak tread grip, which is easy to lead to plane side sliding or rush out of the runway in rain and snow weather, which cannot meet the actual needs; Aviation tires also have a mandatory function of conducting static electricity. The friction between the tires and the ground during the take-off and landing of the aircraft, the friction between the surface of the body and the air in flight, and the electromagnetic mutual inductance generated by the operation of the aircraft control electrical equipment will generate a large amount of charge accumulation and generate static electricity. High-potential static electricity can cause electrical equipment to work instability, and even breakdown and burn electronic components, which is prone to fire when refueling. Static electricity is also very harmful to the body of drivers and passengers, especially under the premise of continuous improvement of the level of intelligence, static electricity hazards can cause serious safety accidents. In order to ensure the normal operation of aircraft equipment, especially electronic equipment, and protect the drivers and passengers from static electricity, it is stipulated that the static electricity of the aircraft body must be discharged through the tires during the landing and take-off process\cite{128}. Over the years, people have carried out unremitting research to improve the static electricity conducting function of aircraft tires, adding a sufficient amount of conductive carbon black or other conductive media to the raw material formulations of the existing aircraft tires. However, the added static conductive medium enables the aviation tires to meet the electrical conductivity indicators while greatly reducing the strength, toughness, heat dissipation, wear resistance, aging resistance, grip and other physical properties of the aviation tires. This is the "Magic Triangle" effect of material function. Its performance is that in the pursuit of improving a certain physical property index, it has to sacrifice another or more physical property indexes; The static conductive medium used in the existing aviation tires has not only caused a serious imbalance of the comprehensive functional indicators of the aviation tires, but some additives may also have a negative impact on the environment.

Graphene has outstanding mechanical properties: Graphene is a new material with the highest known strength. Its tensile strength is 125GPa, elastic modulus is 1.1TPa, and its strength limit is 42N/m$^2$. Its structure is compact and has good flexibility. For example, packaging bags made of 100 nm graphene film can withstand the pressure of 2T heavy objects; Graphene has outstanding thermal conductivity: the thermal conductivity of graphene at room temperature is 5300W/ (m*k), which is 13 times that of copper; And graphene has good physical and chemical mutual solubility with polymer materials. It can be seen that the powerful and broad graphene is an ideal raw material for aero tires to derive static electricity. It can not only meet the requirements of static electricity, but also can effectively overcome the "magic triangle" effect of material functions, and realize a new material that improves the physical properties of aviation tires.
\section*{Conclusions and prospects}
Graphene is a contemporary new 2D nano-carbon material, and its excellent properties can be reflected in new aerospace materials. The excellent conductivity of graphene is widely used in supercapacitors, solar cells and lithium batteries. Due to its high toughness and high strength, graphene can be used in sensors comparable to the significant planar value of graphite and other materials with high mechanical strength. In addition, graphene also has broad prospects in aviation fields such as anti-corrosion materials, thermoelectric materials, and lightning protection materials. Because graphene is very difficult to produce on a large scale, graphene can only be used to transform or enhance aircraft, not to build them, but they can solve various problems in the aviation field, such as energy transmission, energy storage, as well as lightning protection and deicing protection materials. They can bring immeasurable influence to the aviation industry. In the future, when we want to develop high-reliability and long-life new aviation materials, we can pay more attention to graphene, a material with great potential.
\section*{Acknowledgement}
This work was funded by the National Natural Science Foundation of China (NSFC, No. 61973046) and the Jilin Provincial Department of Science and Technology Project (No. 20220201051GX).
\section*{Keywords}
\noindent Graphene, Two-dimensional materials, Aviation industry, Application value

\end{document}